\documentclass[aps,pre,reprint,superscriptaddress,floatfix]{revtex4-2}

\usepackage{color}
\usepackage{amssymb}
\usepackage{amsmath}
\usepackage[bottom]{footmisc}
\usepackage{mathtools}
\usepackage{acronym}
\usepackage{nicefrac}
\usepackage{graphicx}
\usepackage{mwe}
\usepackage[caption=false,font=footnotesize]{subfig}
\usepackage[dvipsnames]{xcolor}
\usepackage[colorlinks=true]{hyperref}
\usepackage{tikz}
\usepackage{pgfplots}
\hypersetup{
	linkcolor = {blue},
	anchorcolor = blue,
	citecolor = blue,
	filecolor = blue,
	urlcolor = blue
}

\newcommand\mb[1]{\boldsymbol{#1}}

\newcommand\Cov[2]{\mathrm{cov}(#1,#2)}
\newcommand\V[1]{\mathrm{var}\left(#1\right)}

\newcommand{\ssep}{\,;\,}

\newacro{FPR}{false-positive rate}
\newacro{AR}{autoregressive}
\newacro{MA}{moving-average}
\newacro{TPR}{true-positive rate}
\newacro{LR}{log-likelihood ratio}
\newacro{FIR}{finite-impulse response}
\newacro{IIR}{infinite-impulse response}

\newcommand{\fig}[1]{Fig.~\ref{fig:#1}}
\newcommand{\secRef}[1]{Sec.~\ref{sec:#1}}

\newcommand{\app}[1]{Appendix~\ref{app:#1}}

\begin{document}
	
	
	\title{Assessing the Significance of Directed and Multivariate Measures of Linear Dependence Between Time Series}
	
	
	\author{Oliver M. Cliff}
	\email{oliver.cliff@sydney.edu.au}
	\affiliation{Centre for Complex Systems, The University of Sydney, Sydney NSW 2006, Australia}
	\affiliation{School of Physics, The University of Sydney, Sydney NSW 2006, Australia}
	\affiliation{Faculty of Engineering, The University of Sydney, Sydney NSW 2006, Australia}
	\author{Leonardo Novelli}
	\affiliation{Centre for Complex Systems, The University of Sydney, Sydney NSW 2006, Australia}
	\affiliation{Faculty of Engineering, The University of Sydney, Sydney NSW 2006, Australia}
	\author{Ben D. Fulcher}
	\affiliation{Centre for Complex Systems, The University of Sydney, Sydney NSW 2006, Australia}
	\affiliation{School of Physics, The University of Sydney, Sydney NSW 2006, Australia}
	\author{James M. Shine}
	\affiliation{Centre for Complex Systems, The University of Sydney, Sydney NSW 2006, Australia}
	\affiliation{Brain and Mind Centre, School of Medical Sciences, The University of Sydney,\\Sydney NSW 2006, Australia}
	\author{Joseph T. Lizier}
	\affiliation{Centre for Complex Systems, The University of Sydney, Sydney NSW 2006, Australia}
	\affiliation{Faculty of Engineering, The University of Sydney, Sydney NSW 2006, Australia}
	
	
	
	\begin{abstract}
		
		Inferring linear dependence between time series is central to our understanding of natural and artificial systems.
		Unfortunately, the hypothesis tests that are used to determine statistically significant directed or multivariate relationships from time-series data often yield spurious associations (Type I errors) or omit causal relationships (Type II errors).
		This is due to the autocorrelation present in the analysed time series---a property that is ubiquitous across diverse applications, from brain dynamics to climate change.
		Here we show that, for limited data, this issue cannot be mediated by fitting a time-series model alone (e.g., in Granger causality or prewhitening approaches), and instead that the degrees of freedom in statistical tests should be altered to account for the effective sample size induced by cross-correlations in the observations.
		This insight enabled us to derive modified hypothesis tests for any multivariate correlation-based measures of linear dependence between covariance-stationary time series, including Granger causality and mutual information with Gaussian marginals.
		We use both numerical simulations (generated by autoregressive models and digital filtering) as well as recorded fMRI-neuroimaging data to show that our tests are unbiased for a variety of stationary time series.
		Our experiments demonstrate that the commonly used $F$- and $\chi^2$-tests can induce significant false-positive rates of up to 100\% for both measures, with and without prewhitening of the signals.
		These findings suggest that many dependencies reported in the scientific literature may have been, and may continue to be, spuriously reported or missed if modified hypothesis tests are not used when analysing time series.
	\end{abstract}
	
	
	\maketitle

\section{Introduction}
	\label{sec:introduction}

	Linear dependence measures such as Pearson correlation, canonical correlation analysis, and Granger causality are used in a broad range of scientific domains to investigate the complex relationships in both natural and artificial processes. Despite their widespread use, concerns have been raised about the hypothesis tests typically used to assess the statistical significance of such measures from time series~\cite{davey2013filtering,afyouni2019effective,florin2010effect,barnett2011behaviour,seth2010matlab,robinson1973generalized}. Specifically, the presence of autocorrelation in a signal---one of two defining properties of a stationary time series~\cite{brockwell1991time,reinsel2003elements}---has been known to bias statistics since the beginning of time-series analysis~\cite{yule1926we}.
	If left unaccounted, this bias yields a greater number of both spurious correlations and missed causalities (Type I and Type II errors) due to size and power distortions of the hypothesis tests.
	With the recent findings~\cite{davey2013filtering,afyouni2019effective,florin2010effect,barnett2011behaviour,seth2010matlab,robinson1973generalized} suggesting that existing techniques do not adequately address autocorrelation, the accuracy of many reported results across the empirical sciences may be called into question.
	
	The notion that autocorrelation affects the sampling distribution of time-series properties has a long history in statistics, with research often focusing on the relationship between two univariate processes.
	Seminal work by Bartlett~\cite{bartlett1935some,bartlett1946theoretical} revealed that autocorrelation can distort the degrees of freedom available to compute statistics such as Pearson correlation coefficients.
	In practical terms, this induces an ``effective sample size'', where the effective number of independent samples used in computing an estimate is different to the actual length of the dataset.
	Two opposing strategies have been proposed for handling autocorrelation: remove the \ac{AR} components of the time series before computing statistics, or modify the hypothesis tests that assess the distorted measurements.
	The former approach, known as prewhitening, involves filtering the time series in order to render the residuals serially independent~\cite{cryer2008time}.
	Prewhitening is known to have many issues, such as reducing the size and power properties of hypothesis tests both in theory~\cite{sul2005prewhitening} and in practice, with simulated~\cite{bayazit2007prewhiten} and recorded~\cite{olszowy2019accurate,yue2002applicability} time-series data in a variety of domains.
	In contrast, the notion of modifying hypothesis tests remains relatively underused in practice, more often found in applications involving short time series and high autocorrelation, where the statistical bias of measures is most pronouced (with or without prewhitening), e.g., in fMRI-based neuroimaging~\cite{arbabshirani2014impact,davey2013filtering,afyouni2019effective}, as well as environmental and ecological studies~\cite{bence1995analysis,macias2012persistence}.
	Indeed, it was not until recently that the efficacy of a modified $z$-test for correlation analysis was demonstrated successfully on fMRI signals~\cite{afyouni2019effective}, which have been widely characterised using correlation coefficients~\cite{friston2011functional}.
	Nevertheless, the theory of autocorrelation on the undirected relationship between bivariate time series is now well developed. However, the extension of this theory to multiple time series, and to directed relationships, remains incomplete.

	Motivated to study directed dependencies in economics, Granger~\cite{granger1969investigating} introduced a measure of causal influence between \ac{AR} models nearly 60 years ago.
	Since then, it has become exceedingly popular, exemplified by more than 100\,000 works indexed by Google Scholar that contain the phrase ``Granger causality'' (as of June, 2020).
	This impact is reflected in the measure's ubiquity in the scientific community beyond its origins in econometrics, generating highly influential results on phenomena ranging from brain dynamics~\cite{roebroeck2005mapping,Liao2011,friston2013analysing} to climate change~\cite{kaufmann1997evidence,zhang2011causality} and political relationships~\cite{freeman1983granger,reuveny1996international}.
	Granger causality controls for the confounding past of a process through linear regression, building statistics and hypothesis tests via residuals rather than the original process.
	However, researchers are becoming aware that certain preprocessing techniques that increase autocorrelation, such as filtering, raise the \ac{FPR} of Granger causality tests when using the well-established $\chi^2$- and $F$-distributions~\cite{florin2010effect,barnett2011behaviour,seth2010matlab}.
	Even though these empirical studies have demonstrated that established Granger causality tests have distorted size and power properties (exhibiting Type I and II errors), it has remained unclear as to why and how to correct them.
	In this paper, we illustrate that these errors are due to an inflated variance of the null distribution as a function of autocorrelation remaining in the residuals, in the same way that bivariate correlation is affected.

	In order to unify Bartlett's earlier investigations on correlation coefficients (under autocorrelation) with more complex measures such as Granger causality, we must expand the former body of work to account for multivariate relationships.
	One such multivariate generalisation of Pearson correlation is referred to as Wilks' criterion~\cite{wilks1932certain}, which quantifies the relationship between multiple sets of variables, and is $\Lambda$-distributed for independent observations~\cite{pham2008exact}.
	In particular, we are interested in a special case of Wilks' criterion popularised by Hotelling~\cite{hotelling1936relations} that focuses on two sets of variables, referred to as canonical correlation analysis.
	It was later established that, like Pearson correlation, estimates of canonical correlations are inefficient under autocorrelation~\cite{robinson1973generalized}, introducing Type I and II errors under hypothesis tests that assume independence (such as the $\Lambda$-distribution).
	Instead of deriving hypothesis tests directly for Wilks' criterion or canonical correlations, here we use the equivalent information-theoretic formulation.
	Information theory's general applicability arises in simply requiring a probability distribution that can be either parametric or non-parametric~\cite{mackay2003information,cover2012elements}.
	When this probability distribution is modelled as a multivariate Gaussian, canonical correlation analysis and information theory overlap because mutual information can be decomposed into sums involving canonical variables~\cite{kay1992feature}.
	Moreover, Granger causality is now understood as a special case of conditional mutual information, known as transfer entropy~\cite{barnett2009granger,schreiber2000measuring,bossomaier2016introduction}.
	While this unification provides an elegant perspective, there remains a clear divide between the theoretical foundations of Bartlett (and others~\cite{bayley1946effective,clifford1989assessing,roy1989asymptotic}) and the large family of multivariate linear dependence measures that information theory provides.

	In this work we bridge this gap by leveraging the concept of the effective sample size to derive hypothesis tests for any correlation-based measure of linear dependence between covariance-stationary time series.
	This comprises a large family of well-known statistics based on ratios of generalised variance---such as Granger causality and mutual information---that we introduce in \secRef{preliminaries}.
	To achieve this, we first provide the one-tailed and two-tailed tests for the sample partial correlations between two univariate processes under autocorrelation in \secRef{exact-tests-partial-correlation}.
	Although this result is important in its own right, in this work we primarily leverage it to construct the tests for more advanced inference procedures with multivariate and directed models of observed~dynamics.
	Following this, we introduce the modified $\Lambda$-test (in \secRef{lambda-tests}), which we show is suitable for assessing the significance of any linear dependence measure that can be expressed as a ratio of generalised variances.
	Specifically, in \secRef{exact-tests-mutual-information} we use the two-tailed test to derive hypothesis tests for conditional mutual information estimates between bivariate time-series data.
	We then use the chain rule for mutual information to extend this result to multivariate time-series data. 
	Finally, since Granger causality can be expressed as a conditional mutual information, in \secRef{exact-tests-granger-causality} we extend our results further to derive Granger causality tests for both bivariate and multivariate time-series datasets.
	More broadly, the modified $\Lambda$-test can be used for any measure that can be expressed in terms of conditional mutual information (or, equivalently, Wilks' criterion or partial correlation), e.g., canonical correlations and partial autocorrelation~\cite{reinsel2003elements,brockwell1991time} or information-theoretic measures (for linear-Gaussian processes) such as predictive information~\cite{bialek2001complexity,crutchfield2003regularities} and active information storage~\cite{lizier2012local}.

	Using numerical simulations throughout \secRef{numerical-simulations}, we validate the modified $\Lambda$-test and characterise the effect of autocorrelation on both the $\chi^2$- and $F$-test. Our experiments involve generating samples from two first-order independent \ac{AR} models and iteratively filtering the output signal such that the autocorrelation is increased for both time series; this simulates empirical analysis in practise, and allows for the process parameters to be modified while ensuring that the null hypothesis (of no inter-process dependence) is not violated.
	We perform these experiments for mutual information and Granger causality in their unconditional, conditional, and multivariate forms. Our results generally agree with the hypotheses that the \ac{FPR} of $F$- and $\chi^2$-tests can be inflated by either increasing the autocorrelation (through filtering) or, for the $\chi^2$-test, the number of conditionals (through increasing the dimension of mutual information or the history length of Granger causality).
	These experiments mirror empirical applications where digital filtering is often used in preprocessing for many purposes, such as handling nonstationary effects, which inadvertently increases autocorrelation and therefore the FPRs of unmodified tests.
	Given minimally sufficient effective samples, however, we confirm that the modified $\Lambda$-tests remain unbiased for all scenarios.
	We thus show that, in contrast, the size (Type I errors) and power (Type II errors) of $F$- or $\chi^2$-tests are arbitrarily low for a large class of multivariate linear dependence measures (approximately zero in certain instances) and overwhelmingly depends on the parameters of the underlying independent processes.
	We further demonstrate that the common approach of prewhitening a signal (in order to remove the effect of autocorrelation) does not suffice to control the \ac{FPR} in almost all cases.
	Finally, by using a well-known brain-imaging dataset from the Human Connectome Project~\cite{van2012human}, we verify that our previous numerical simulations yield comparable results to experiments on commonly used datasets.
	For these experiments, the $\chi^2$-tests of mutual information and Granger causality yield concerningly high \acp{FPR} of over 80\% and 65\% for a nominal significance of 5\%---a 16- and a 13-fold increase---whereas our exact tests maintain the ideal \ac{FPR} for all experiments.
	Open-source MATLAB code is made available to allow users to perform correct hypothesis testing for all dependence measures, as well as the above experiments, at: \url{https://github.com/olivercliff/exact-linear-dependence}.
	
	Our theoretical and empirical findings suggest that this work presents the first statistically sound approach for testing the linear dependence between multivariate time-series data. Given that approaches such as prewhitening and Granger causality are specifically designed to account for autocorrelation, we conjecture that autocorrelation-induced statistical errors caused by $F$- or $\chi^2$-tests (and others) may be even more prevalent in prior publications than previously suggested by a number of authors~\cite{barnett2011behaviour,davey2013filtering,florin2010effect}. 
	In particular, our case study of brain-imaging data is concerning, because the neuroscience community employs techniques such as correlation, mutual information, and Granger causality in order to infer pairwise dependence (known as ``functional connectivity'').
	Implementation of our approach will enable correct inference of linear relationships within complex systems across myriad scientific applications.

\section{Measures of linear dependence}
\label{sec:preliminaries}

    In this work, we focus on multivariate signals,
	\begin{equation}
		\left\{ Z_{1}(t),\ldots,Z_{m}(t) \right\}, \hspace{10px} t = 0, \pm 1, \pm 2, \ldots,
	\end{equation}
	that is, a collection of $m$ series sampled at equally spaced time intervals. Writing
	\begin{equation} \label{eq:zt}
		\mb{Z}(t) = (Z_{1}(t), \ldots, Z_{m}(t))',
	\end{equation}
	we shall refer to the $m$ series as an $m$-dimensional vector of multiple time series such that $\mb{Z}(t) \in \mathbb{R}^m$.
	For the purposes of inferring linear dependence, $\mb{Z}$ is partitioned into one $k$-variate and one $l$-variate subprocess~\cite{geweke1982measurement}:
	\begin{equation} \label{eq:partition}
	\mb{Z} = \begin{bmatrix}
	\mb{X} \\ \mb{Y}
	\end{bmatrix},
	\end{equation}
	reflecting an interest in the relationship between $\mb{X}$ and $\mb{Y}$.
	The linear dependence of $\mb{X}$ on $\mb{Y}$ (or vice versa) is measured by a scalar value that quantifies how much the outcomes of $\mb{Y}$ reduce uncertainty over outcomes of $\mb{X}$.
	Theoretically, in the absence of a linear relationship between $\mb{X}$ and $\mb{Y}$ (the null hypothesis, $\mathcal{H}_0$) the reduction of uncertainty is exactly zero, meaning that $\mb{Y}$ does not linearly predict $\mb{X}$ at all.
	In practice, however, we only have access to a finite-length dataset with $T$ observations over which to compute the measures, introducing a variation in statistical estimates and manifesting as non-zero values in the case of no relationship.
	Here, we present this dataset as an $m\times T$ matrix $\mb{z}$ of consecutive real-valued samples $\mb{z}(t) \in \mathbb{R}^m$ of the process $\mb{Z}$ (again, this is partitioned into submatrices $\mb{x}$ and $\mb{y}$).
	To this end, the aim of linear-dependence tests is to infer whether there is a statistical dependence between $\mb{X}$ and $\mb{Y}$ based on the sample paths $\mb{x}$ and $\mb{y}$ alone.

	We make the typical assumption that the underlying system, $\mb{Z}$, is a second-order stationary, purely non-deterministic process~\cite{box2015time,brockwell1991time,reinsel2003elements}.
	An important consequence of covariance-stationarity is that the time series may be represented, after appropriate mean removal and differencing~\footnote{The most general form is the ARIMA model, where the integrative (I) component accounts for non-stationary time series. Here we assume that appropriate differencing and mean removal has already been performed in order to remove any integration of the time series.}, by the ARMA model:
    \begin{equation} \label{eq:arma}
        \mb{Z}(t) = \mb{a}(t) + \sum_{u=1}^p \mb{\Phi}(u) \mb{Z}(t-u) + \sum_{u=1}^q \mb{\Theta}(u) \mb{a}(t-u),
    \end{equation}
    where $\mb{\Phi}$ and $\mb{\Theta}$ are vectors of autoregressive (AR) and \ac{MA} parameters, and $\mb{a}(t)$ is uncorrelated noise (the innovation process).
    We further assume that the noise is Gaussian, $\mb{a}(t)\sim \mathcal{N}(\mb{0},\mb{\Sigma})$ for some arbitrary noise covariance $\mb{\Sigma}$, meaning that $\mb{Z}$ is a linear-Gaussian process.

	\subsection{Cross-correlation and autocorrelation}
	
	For covariance-stationary time series, the relationship between $Z_i(t)$ and $Z_j(t+u)$ depends only on the difference in times $t$ and $t+u$ of the observation but not on $t$ itself.
	Once the mean has been removed, such processes are fully defined by their cross-correlation,
	\begin{equation} \label{eq:corr}
		\rho_{ij}(u) = \frac{\gamma_{ij}(u)}{\sqrt{\gamma_{ii}(0) \gamma_{jj}(0)}},
	\end{equation}
	with $\gamma_{ij}(u) = \Cov{Z_i(t)}{Z_j(t+u)}$ the cross-covariance between $Z_i(t)$ and $Z_j(t+u)$.
	If $\rho_{ii}(u) \neq 0$ for any $u>0$, then the univariate process $Z_i$ exhibits autocorrelation, and the collection of $\rho_{ii}(u)$ for $u = 0, \pm 1, \pm2,\ldots$ is generally called the autocorrelation function of $Z_i$.
	The sample cross-correlation coefficients are computed from time-series data $\mb{z}$ as
	\begin{equation} \label{eq:inf-corr}
		r_{ij}(u) = \frac{c_{ij}(u)}{\sqrt{c_{ii}(0) c_{jj}(0) }},
	\end{equation}
	with $c_{ij}(u) = N^{-1} \sum_{t=1}^T z_{i}(t) z_{j}(t+u)$ where $N=T-1$ as an unbiased estimate of the sample cross-covariance.
	
	The first linear dependence measure we discuss is Pearson's product-moment correlation coefficient.
	For bivariate ($m=2$) processes, we shall write $\mb{X} = X$ and $\mb{Y} = Y$ and denote the cross-correlation between these variables (Eq.~\eqref{eq:corr}) as $\rho_{XY}(u)$.
	Pearson's correlation coefficient is the lag-zero cross-correlation $\rho_{XY} = \rho_{XY}(0)$, and quantifies the (symmetric) association between paired observations of $X$ and $Y$.
	The sample correlation coefficient is then given by
	\begin{equation}
	    r_{xy} = \frac{c_{xy}}{c_{xx} c_{yy}},
	\end{equation}
	where $c_{xy} = c_{xy}(0)$ is the sample covariance, and $c_{xx}=c_{xx}(0)$ and $c_{yy}=c_{yy}(0)$ are sample variances~\cite{clifford1989assessing}.
	
	In order to assess the statistical significance of linear dependence measures, such as the sample correlation coefficient, we must be able to compute their variance, i.e., $\sigma_r^2(x,y) = \V{r_{xy}}$. For independent but autocorrelated stationary processes, the variance of the sample correlation coefficient can be estimated (to the first order) as~\cite{clifford1989assessing,afyouni2019effective,bayley1946effective}
	\begin{equation} \label{eq:bartletts-var}
	    \hat{\sigma}_r^2(x,y) \approx T^{-1} \left( 1 + 2 \, \sum_{u=1}^{T-1} \frac{T-u}{T} r_{xx}(u) r_{yy}(u) \right),
	\end{equation}
	where $r_{xx}(u)$ and $r_{yy}(u)$ are the lag $u$ sample autocorrelations~\footnote{The problem of recovering the autocorrelation functions from data, and thus the effective sample size, is nontrivial and has resulted in procedures such as tapering to handle noisy estimates~\cite{chatfield2003analysis,afyouni2019effective}; these notions are covered briefly in Appendix~\ref{sec:apx-mv-bartletts}.}.
	Although we refer to Eq.~\eqref{eq:bartletts-var} as Bartlett's formula, this first-order approximation is due to Clifford et al.~\cite{clifford1989assessing}, who presented the variance estimator for spatial autocorrelation and an estimate of the effective number of independent samples for correlation coefficients:
	\begin{equation} \label{eq:bartletts}
		\hat{\eta}(x,y) = 1 + \hat{\sigma}_r^{-2}(x,y).
	\end{equation}
	An important consequence of Eqs.~\eqref{eq:bartletts-var} and~\eqref{eq:bartletts} is that hypothesis tests, such as Student's $t$-test, should have degrees of freedom corresponding to the effective sample size of the analysed time series (i.e., the effective degree of freedom~\cite{afyouni2019effective,bartlett1946theoretical}), rather than the original sample size~\cite{clifford1989assessing}. That is, if both $X$ and $Y$ are autocorrelated, then the null distribution for $r_{xy}$ follows a modified Student's $t$-test:
		\begin{equation} \label{eq:t-test-bart}
			r_{xy} \sqrt{\frac{\hat{\eta}(x,y)-2}{1-r_{xy}^2}} \sim \, t(\hat{\eta}(x,y)-2),
		\end{equation}
		where $\hat{\eta}(x,y)-2$ is the (estimated) effective degrees of freedom.
	An examination of this formula reveals that, when both $x$ and $y$ are positively autocorrelated, then there are, effectively, fewer independent observations than in the original dataset ($\hat{\eta} < T$); if only one process is negatively autocorrelated, then there appear to be more independent observations than in the original dataset ($\hat{\eta}>T$)~\cite{clifford1989assessing}.
	Consequently, when the modified degree of freedom in Eq.~\eqref{eq:t-test-bart} is neglected, inference procedures can either spuriously identify association (produce Type I errors) when $\hat{\eta} < T$ or miss actual correlations (Type II errors) when $\hat{\eta} > T$.
		
	If either one (or both) of $x$ or $y$ are serially independent, then the sample correlation coefficients $r_{xy}$ can be tested against Student's $t$-distribution with degrees of freedom $T-2$.
	Thus, the textbook approach for minimising the deleterious effects of autocorrelation is to whiten one of the time series by filtering any \ac{AR} components (referred to as prewhitening).
	The idea is that, by filtering any \ac{AR} components, the residuals become uncorrelated and so statistical tests that have been developed for independent variables can now be used without modifying the degree of freedom.
	In \secRef{prewhitening}, we discuss this approach in more detail, showing that linear-dependence tests applied to signals that have been ``whitened'' in this way still exhibit significant statistical bias (in some cases worse than without prewhitening).

	\subsection{Partial correlation}

		Partial correlation $\rho_{XY\cdot \mb{W}}$ measures the association between $X$ and $Y$, whilst controlling for any concomitant effect of another $c$-variate process $\mb{W}$~\cite{anderson2001permutation,chatfield2003analysis}.
		Partial correlation is estimated by, first, computing the residual processes:
		\begin{align}
			e_{x\mid\mb{w}} &= x - \hat{x}(\mb{w}) \label{eq:residual-x} \\
			e_{y\mid\mb{w}} &= y - \hat{y}(\mb{w}), \label{eq:residual-y}
		\end{align}
		where $\hat{x}(\mb{w})$ denotes the linear prediction of $x$ from $\mb{w}$ via ordinary least squares.
		Then, an appropriate test statistic for the null hypothesis $\mathcal{H}_0:\rho_{XY\cdot \mb{W}} = 0$ of no relation between $X$ and $Y$, above any relationship with $\mb{W}$, is the sample partial correlation:
		\begin{equation} \label{eq:inf-parcorr}
			r_{xy\cdot \mb{w}} = \frac{\sum_t e_{x \mid \mb{w}}(t) \, e_{y \mid \mb{w}}(t) }{\sqrt{\sum_t e^2_{x \mid \mb{w}}(t)} \sqrt{\sum_t e^2_{y \mid \mb{w}}(t)} }.
		\end{equation}
		By contrasting the formulas for sample partial correlation (Eq.~\eqref{eq:inf-parcorr}) with sample cross-correlation (Eq.~\eqref{eq:inf-corr}), it is evident that the former is equivalent to the bivariate correlation between the residuals, i.e., $r_{xy\cdot \mb{w}} = r_{e_{x\mid \mb{w}} e_{y\mid\mb{w}}}$.
		
		Unlike Pearson correlation, there is a dearth of research into the null distribution of partial correlation coefficients for autocorrelated time-series data.
		As such, our first theoretical contribution (in \secRef{exact-tests-partial-correlation}) is a derivation for the null distribution of sample partial correlations~\eqref{eq:inf-parcorr} under autocorrelation, i.e., extending the modified $t$-test (for bivariate correlation~\eqref{eq:t-test-bart}) to facilitate residual processes.
		
		\subsection{Wilks' criterion and canonical correlations}

		Relating two or more sets of variables is achieved similarly to partial correlation~\eqref{eq:inf-parcorr}, with the exception that the generalised variance is used, rather than the conditional variance~\cite{wilks1932certain,hotelling1936relations}.
		Consider the relationship between the $k$-variate process $\mb{X}$ and the $l$-variate process $\mb{Y}$, in the context of a $c$-variate concomitant process $\mb{W}$.
		To measure their dependence, we use the same procedure as for univariate processes, except now the residuals $\mb{e_{x\mid w}}$ and $\mb{e_{y\mid w}}$ (from Eqs.~\eqref{eq:residual-x} and~\eqref{eq:residual-y}) are multivariate, making the sample covariance $\mb{s_{xy\mid w}}$ an $m\times m$ matrix, rather than a scalar value.
		The generalised sample variance is the determinant of these sample covariances $\lvert \mb{s_{xy\mid w}} \rvert$, and can be used to form a special case of (residual) Wilks' criterion~\cite{wilks1932certain}:
		\begin{equation} \label{eq:wilks}
			\frac{\lvert \mb{s_{xy\vert \mb{w}}} \rvert}{ \lvert \mb{s_{x \vert \mb{w}}} \rvert \lvert \mb{s_{y \vert \mb{w}}} \rvert }.
		\end{equation}
		Although in general Wilks' criterion facilitates any number of partitions of $\mb{Z}$, we will restrict our attention to two partitions (referring to the special case in Eq.~\eqref{eq:wilks} as Wilks' criterion when the meaning is clear).
		The null distribution of the ratio of independent generalised variances~\eqref{eq:wilks} is known as Wilks' $\Lambda$-distribution, with its exact analytic form derived by a number of authors on the basis of no cross-correlation and under the hypothesis that each variable within $\mb{X}$, $\mb{Y}$, or $\mb{W}$ exhibit no autocorrelation~\cite{pham2008exact,mathai1971exact}.
		Hotelling~\cite{hotelling1936relations} extensively studied the case of Wilks' criterion with two sets of variables, showing invariance under any internal linear transformation of these sets and a decomposition into canonical correlations with an asymptotic ($\chi^2$) null distribution.
		Much like their univariate counterparts, however, Hotelling's canonical correlations have been shown to be inefficient under autocorrelation~\cite{robinson1973generalized}.
		Consequently, neither approach is suitable for inferring linear-dependence between the majority of time-series data due to the ubiquity of autocorrelation.
		In \secRef{lambda-tests}, we address this issue by providing the null distribution to be used in the presence of autocorrelation (and of course when cross-correlations are present amongst any two variables, i.e., $X_i(t)$ and $X_j(t-u)$ may covary for any $i$, $j$, $t$, or $u$).
		An application that is of particular interest is mutual information, which is equivalent to Wilks' criterion~\eqref{eq:wilks} for Gaussian marginals.

	\subsection{Mutual information}

		Mutual information $\mathcal{I}_{\mb{X};\mb{Y} \vert \mb{W}}$ is a fundamental concept in information theory---and a building block of many other measures---that quantifies the amount of information about a process $\mb{X}$ obtained by observing another process $\mb{Y}$ (potentially in the context of a third process $\mb{W}$, making it a conditional mutual information)~\cite{cover2012elements}.
		In general, information theory facilitates multivariate analysis by simply requiring well-defined probability distributions that can be either parametric or non-parametric.
		When these are normally distributed, mutual information takes a form that is equivalent to Wilks' criterion~\cite{cover2012elements,brillinger2004some}:
		\begin{equation} \label{eq:cmi}
			\hat{\mathcal{I}}_{\mb{x};\mb{y} \mid \mb{w}} = -\frac{1}{2} \log \left( \frac{\lvert \mb{s_{xy\vert w}} \rvert}{ \lvert \mb{s_{x \vert w}} \rvert \lvert \mb{s_{y \vert w}} \rvert } \right).
		\end{equation}
		This formula is asymptotically equivalent to the nested \ac{LR} of two models~\cite{barnett2012transfer}, and thus we can use a null distribution also provided by Wilks~\cite{wilks1938large}. Following Wilks' theorem~\cite{wilks1938large}, under the null hypothesis $\mathcal{H}_0 : \mathcal{I}_{\mb{X};\mb{Y} \mid \mb{W}}=0$ and with normally distributed marginals, mutual information estimates are asymptotically chi-square distributed~\cite{brillinger2004some,barnett2012transfer,barnett2009granger},
		\begin{equation} \label{eq:dist-cmi-asymp}
			2 T\, \hat{\mathcal{I}}_{\mb{x};\mb{y} \mid \mb{w}} \sim  \chi^2(kl).
		\end{equation}
		For limited data, a more precise null distribution can be derived from the standard $F$-test, albeit for the more specific case of no autocorrelation and with one of the processes being univariate (see Eq.~\eqref{eq:f-test-lr} and surrounding discussion in \app{f-test}).
		That is, the mutual information between an i.i.d.\ variable $X$ and an $l$-variate $\mb{Y}$, in the context of the concomitant $\mb{W}$, is, under the null hypothesis $\mathcal{H}_0 : \mathcal{I}_{X;\mb{Y} \mid \mb{W}}=0$,
		\begin{equation} \label{eq:dist-cmi-f}
			\frac{T-(l+c+1)}{l}[ \exp{(2 \hat{\mathcal{I}}_{x;\mb{y} \mid \mb{w}})} - 1 ] \sim F(l, T-(l+c+1)).
		\end{equation}
		The same arguments in Eqs.~\eqref{eq:cmi}--\eqref{eq:dist-cmi-f} hold for unconditional mutual information $\hat{\mathcal{I}}_{\mb{x};\mb{y}}$ by setting $\mb{w} = \emptyset$ (and $c = 0$).
		
		Although we found no discussion on autocorrelation-induced biases of mutual information in literature, statistical tests will clearly be incorrect if autocorrelation is not taken into account.
		This is evident from the well-known result that mutual information reduces to a function of sample correlation coefficients when the dependent processes are univariate~\cite{davey2013equivalence}:
	    \begin{equation} \label{eq:inf-cmi}
	        2\,\hat{\mathcal{I}}_{x;y\mid \mb{w}} = -\log(1-r_{xy \cdot \mb{w}}^2).
	    \end{equation}
		Moreover, it is clear by observing the equivalence between Eqs.~\eqref{eq:wilks} and~\eqref{eq:cmi}, noting that mutual information estimates can be decomposed into sums involving canonical variables~\cite{brillinger2004some}.
		By deriving the exact hypothesis tests for mutual information in \secRef{exact-tests-mutual-information}, we provide a critical component of general-purpose techniques for measuring undirected relationships between sets of variables.
		However, mutual information was not originally intended to measure autocorrelated time-series dependencies, nor does it naturally model directed dependencies---this is the intended purpose of Granger causality.

	\subsection{Granger causality}

		Granger causality was explicitly designed to capture one-way dependence between stochastic processes by taking into account the confounding influence of their past (i.e., the autocorrelation).
		By considering $\mb{X}$ as a target (predictee) process and $\mb{Y}$ as a source (predictor) process, Granger causality $\mathcal{F}_{\mb{Y} \rightarrow \mb{X} \mid \mb{W}}$ explicitly aims to measure the causality (predictability) in $\mb{Y}$ about $\mb{X}$ in context of the relevant history of $\mb{X}$ (and, potentially, a concomitant process $\mb{W}$).
		Of course, the use of the term ``causality'' here refers to Wiener's definition (as a model of dependence based on prediction) rather than Pearl's (a mechanistic causal-effect that can only be inferred using interventions); see~\cite{lizier2010differentiating} for a differentiation of these concepts.
		
		The main assumption underlying Granger causality is that both $\mb{X}$ and $\mb{Y}$ are (vector) \ac{AR} processes~\cite{geweke1982measurement,granger1969investigating}.
		That is, we assume that $\mb{X}(t)$ and $\mb{Y}(t)$ are causally dependent on the following states:
		\begin{equation}
			\mb{X}^{(p)}(t) = \begin{bmatrix} \mb{X}(t-1) \\ \vdots \\ \mb{X}(t-p) \end{bmatrix}, \hspace{2mm} \mb{Y}^{(q)}(t) = \begin{bmatrix} \mb{Y}(t-1) \\ \vdots \\ \mb{Y}(t-q) \end{bmatrix}.
		\end{equation}
		Under this assumption, the (directed) influence from $\mb{Y}$ to $\mb{X}$ is quantified by conditional mutual information~\cite{barnett2009granger}:
		\begin{equation} \label{eq:mvgc}
			\mathcal{F}_{\mb{Y} \rightarrow \mb{X} \mid \mb{W}}(p,q) = 2\,\mathcal{I}_{\mb{X} ; \mb{Y}^{(q)} \vert \mb{X}^{(p)} \mb{W} }.
		\end{equation}
		Following Eq.~\eqref{eq:cmi}, this measure can be estimated as a log-ratio of generalised variances~\cite{geweke1982measurement}:
		\begin{equation} \label{eq:est-gc}
			\hat{\mathcal{F}}_{\mb{y}\to \mb{x} \mid \mb{w}}(p,q) = \log \Bigg( \frac{\big\lvert \mb{s}_{\mb{x} \mid \mb{x}^{(p)}\mb{w}} \big\rvert}{\big\lvert \mb{s}_{\mb{x} \mid \mb{x}^{(p)} \mb{y}^{(q)}\mb{w}} \big\rvert} \Bigg).
		\end{equation}
		Note that, except for the rather narrow case of $k=q=1$, Granger causality is a multivariate measure (using generalised variances rather than conditional variances).
		
		The \ac{AR} orders, $p$ and $q$, of each process are typically determined by statistical tests such as partial autocorrelation~\cite{reinsel2003elements}, Burg's method~\cite{de1996yule}, the Akaike or Bayesian information criterion (AIC or BIC), cross-validation~\cite{barnett2011behaviour}, or active information storage~\cite{lizier2012local,garland2016}~\footnote{In general, the variables representing the past need not be a sequence of consecutive temporal indices, nor have the same history length $p$ for each dimension of $\mb{X}$. For instance, one could use a Takens embedding~\cite{Takens1981,cliff2018minimising} or any other statistically significant set of variables~\cite{novelli2019large}.
		However, for simplicity, in this work we follow the vector \ac{AR} model often used in Granger causality analysis (Eq.~\eqref{eq:mv-ar}) and thus use a consecutive sequence of variables as the history of $\mb{X}(t)$.}. 
		In this paper, we use Burg's method to infer the model order due to its efficiency and stability over the Yule-Walker equations~\cite{de1996yule}.
		Further, using results from the main text, we discuss the relationship between partial autocorrelation and active information storage in Appendix~\ref{sec:apx-ais}.
		
		In general, hypothesis tests for Granger causality can be derived from Wilks' theorem~\cite{wilks1938large}. That is, under the null hypothesis $\mathcal{H}_0 : \mathcal{F}_{\mb{Y} \to \mb{X} \mid \mb{W}}(p,q) = 0$, estimates of the Granger causality from $\mb{X}$ to $\mb{Y}$ (Eq.~\eqref{eq:est-gc}) are asymptotically chi-square distributed~\cite{geweke1982measurement}:
		\begin{equation} \label{eq:dist-mvgc-asymp}
		T \, \hat{\mathcal{F}}_{\mb{y}\to \mb{x} \mid \mb{w}}(p,q) \sim \chi^2(klq).
		\end{equation}
		Alternatively, a finite-sample null distribution is can be used if the predictee process is univariate ($k=1$).
		Referring to Appendix~\ref{apx:hypothesis-tests}, the restricted model has $p+c+1$ parameters, and the unrestricted model has $p+lq+c+1$ parameters.
		Accordingly, we can build the statistic:
		\begin{equation}
			\frac{T-(p+lq+c+1)}{lq} [\exp{(\hat{\mathcal{F}}_{\mb{y} \to x \mid \mb{w}}(p,q))}-1],
		\end{equation}
		which, according to the standard $F$-test (Eq.~\eqref{eq:f-test-lr}), is distributed as
		\begin{equation} \label{eq:dist-gc-f}
			F(lq,T-(p+lq+c+1)).
		\end{equation}
		It should be emphasised, however, that the $F$-test is only suitable for serially independent observations.
		Thus, although the Granger causality measure explicitly accounts for autocorrelation in vector \ac{AR} processes, the established hypothesis tests assume either a sequence of completely independent residuals (the $F$-test) or infinite data (the $\chi^2$-test).
		The null distributions that we provide in Sec.~\ref{sec:exact-tests-granger-causality} overcome both of these issues, providing the first valid finite-sample tests for Granger causality.

	\section{Modified tests for partial correlation}
	
	\label{sec:exact-tests-partial-correlation}


		In this section, we derive one-tailed and two-tailed tests for the null hypothesis, $\mathcal{H}_0 : \rho_{XY\cdot \mb{W}} = 0$, of no partial correlation between two univariate autocorrelated time series $x$ and $y$, given a third (potentially multivariate) process $\mb{w}$. These tests are valid for any covariance-stationary time series $X$, $Y$, and $\mb{W}$ and sample size $T$.

		\subsection{Modified Student's $t$-test for partial correlation}
		
			Recall from Eq.~\eqref{eq:inf-parcorr} that the sample partial correlation is equivalent to the sample correlation between $e_{x \mid \mb{w}}$ and $e_{y \mid \mb{w}}$, i.e., $r_{xy\cdot \mb{w}} = r_{e_{x \mid \mb{w}}e_{y \mid \mb{w}}}$.
			Obtaining the null distribution for sample partial correlations between autocorrelated time series can thus be treated similarly to that of the correlation coefficients (see Eq.~\eqref{eq:t-test-bart}).

			A well-known result is that the sample partial correlation between independent observations is $t$-distributed $t(\nu)$, under the null hypothesis  $\mathcal{H}_0 : \rho_{XY\cdot \mb{W}}=0$, with degrees of freedom $\nu = T - c - 2$ and $c = \dim(\mb{w}(t))$~\cite{wishart1928generalised}.
			As such, the modified statistic and null distribution is:
			\begin{equation} \label{eq:t-test-pc-bart}
				r_{xy\cdot \mb{w}} \sqrt{\frac{\hat{\eta} \! \left(e_{x\mid \mb{w}},e_{y\mid \mb{w}}\right) - c - 2}{1-r_{xy\cdot \mb{w}}^2}} \sim t(\hat{\eta} \! \left(e_{x\mid \mb{w}},e_{y\mid \mb{w}}\right) - c - 2).
			\end{equation}
			Here, the (estimated) effective sample size $\hat{\eta}(e_{x \mid \mb{w}},e_{y\mid \mb{w}})$ is still computed from Eq.~\eqref{eq:bartletts} but with the autocorrelation functions of the residual vectors $e_{x\mid \mb{w}}$ and $e_{y\mid \mb{w}}$, rather than the original sample paths $x$ and $y$.
			Intuitively, this is because $r_{xy\cdot \mb{w}}$ is itself a sample correlation of these residuals, so it is their autocorrelation---not that of the original time series---that directly determines the effective sample size.
			Another crucial addition is that the dimension of the conditional process $c = \dim(\mb{w}(t))$ further reduces the number of degrees of freedom~\cite{wishart1928generalised}, for the same reason as in standard $F$-tests~\cite{barnett2011behaviour}.
			When the residual vectors, $e_{x\mid \mb{w}}$ and $e_{y\mid \mb{w}}$, are (serially) independent, then Eq.~\eqref{eq:t-test-pc-bart} becomes equivalent to the standard Student's $t$-distribution for partial correlation.

    		\subsection{Modified $F$-test for partial correlation}
			The Student's $t$-distribution in Eq.~\eqref{eq:t-test-pc-bart} allows for one-tailed (upper or lower) tests for the partial correlation by using the statistic (the LHS) as an input to the quantile function of the $t$-distribution.
			For two-tailed tests, another common approach is to square the statistic and, subsequently, the null distribution.
			The square of a random variable $Z\sim t(\nu)$ that follows Student's $t$-distribution (with parameter $\nu$) follows an $F$-distribution with parameters $1$ and $\nu$, i.e., $Z^2 \sim F(1,\nu)$. Thus, under the null hypothesis $\mathcal{H}_0: \rho_{XY\cdot \mb{W}} = 0$, the square of the statistic in Eq.~\eqref{eq:t-test-pc-bart} (the LHS) follows an $F$-distribution:
			\begin{equation} \label{eq:dist-r2}
				n_{xy \mid \mb{w}}\,\frac{r_{xy\cdot \mb{w}}^2}{1-r_{xy\cdot \mb{w}}^2} \sim F(1,n_{xy \mid \mb{w}}),
			\end{equation}
			with an effective degree of freedom,
			\begin{equation} \label{eq:ess-corr}
				n_{xy \mid \mb{w}} = \hat{\eta}(e_{x\mid \mb{w}},e_{y\mid \mb{w}})-c-2,
			\end{equation}
			obtained from Eq.~\eqref{eq:bartletts}.
			We refer to the significance test that uses this distribution as the modified $F$-test.
			Note that a form of Eq.~\eqref{eq:dist-r2} without modifying the degree of freedom is commonly used for testing the coefficient of determination; thus this approach could also be used for constructing a finite-sample test of the coefficient of multiple correlation under autocorrelation.

		\section{Modified \texorpdfstring{$\Lambda$}{lambda}-tests}
		
		\label{sec:lambda-tests}
    		
    		Although the modified $t$- and $F$-tests introduced above are suitable for bivariate correlation-based measures, they are not appropriate for multivariate (and thus directed) null tests.
    		Here, we introduce the $\Lambda^*$-distribution, which can be used for hypothesis testing all linear dependence measures throughout this paper.
    		
    		Recall that, for independent $\mb{X}$ and $\mb{Y}$ and $\mb{W}$, Wilks' criterion (Eq.~\eqref{eq:wilks}) is $\Lambda$-distributed.
			The exact form of the $\Lambda$-distribution has been extensively studied, with known relationships to the $F$- and beta-distributions~\cite{pham2008exact,mathai1971exact}.
    		The main purpose of this paper is to derive the finite sample distribution of such statistics under autocorrelation, i.e., where $\mb{X}(t) \not\!\perp\!\!\!\perp \mb{X}(t-u)$ and $\mb{Y}(t) \not\!\perp\!\!\!\perp \mb{Y}(t-v)$ for some $u,v > 0$.
    		
    		As we will show throughout this work, the distribution for Wilks' criterion with two independent but serially correlated processes can be described by products of $\Lambda$-distributed variables with different effective degrees of freedom.
			We denote this distribution as $\Lambda^*(\mb{n})$, with the parameter $\mb{n}=(n_1,\ldots,n_b)'$ comprising the degrees of freedom of each independent $\Lambda$-distribution.
			That is, Wilks' criterion is, under the null hypothesis, $\Lambda^*$-distributed:
    		\begin{equation}
        		\frac{\lvert s_{\mb{xy} \mid \mb{w}} \rvert}{\lvert s_{\mb{x} \mid \mb{w}} \rvert \lvert s_{\mb{y} \mid \mb{w}} \rvert} \sim \Lambda^*(\mb{n}),
			\end{equation}
			where the $\Lambda^*(\mb{n})$ distribution itself can be described by a product of independent $\Lambda$-distributed variables:
			\begin{equation} \label{eq:lambdas}
			    \prod_{i=1}^b L_i, \hspace{3mm} \text{with } L_i \sim \Lambda(n_i,1,1).
			\end{equation}
			Notice that this reduces to the $\Lambda$-distribution for two sets of independent variables~\cite{pham2008exact}, however, with the $\Lambda^*$-distribution we are able to include the effective sample sizes.
			
			Although the null distribution for the product of two independent $\Lambda$-distributed variates is known~\cite{pham2008exact}, deriving the exact distribution for the product of an arbitrary number of $\Lambda$-distributed variates is non-trivial.
			Fortunately, a relationship between the beta-, $F$-, and $\Lambda$-distributions~\cite{pham2008exact,mathai1971exact} allows for simple numerical methods.
			To generate the distribution $\Lambda^*(\mb{n})$, we could sample beta-distributed variables:
			\begin{equation} \label{eq:betas}
		        \prod_{i=1}^b L_i = \prod_{i=1}^b V_i, \hspace{3mm} \text{with } V_i \sim B\! \left(\frac{n_i}{2},\frac{1}{2}\right),
			\end{equation}
			where $B(\alpha,\beta)$ is the beta distribution.
			Equivalently, one could sample independent $F$-distributed variables:
			\begin{equation} \label{eq:fs}
				\prod_{i=1}^b L_i = \prod_{i=1}^b \frac{n_i}{U_i + n_i}, \hspace{3mm} \text{with } U_i \sim F(1,n_i).
			\end{equation}
			In our experiments (and open-source code), we opt to sample independent beta-distributed variables and constructing the $\Lambda^*$-distribution from their product as per Eq.~\eqref{eq:betas}. 
			Throughout this work, we refer to hypothesis tests that use the $\Lambda^*$-distribution and modify the degree of freedom to account for autocorrelation as ``modified $\Lambda$-tests''.
			
			From the relationship between the beta-, $F$- and $\Lambda$-distributions (see Eqs.~\eqref{eq:lambdas}--\eqref{eq:fs}), it is clear that the modified $\Lambda$-test is a generalisation of the modified $F$-test, becoming equivalent for univariate statistics.
			For instance, returning to partial correlation, we have that:
			\begin{equation} \label{eq:stat-l1}
			   \frac{\lvert s_{xy \mid \mb{w}} \rvert}{\lvert s_{x \mid \mb{w}} \rvert \lvert s_{y \mid \mb{w}} \rvert} = 1-r_{xy\cdot \mb{w}}^2 \sim \Lambda^*(n_{xy \mid \mb{w}}),
			\end{equation}
			where $n_{xy \mid \mb{w}}$ is the effective degree of freedom.
			Thus, either the modified $F$-test or the modified $\Lambda$-test could be used for univariate statistics (i.e., ratios of conditional variances).

			We can now derive explicit hypothesis tests for common directed and multivariate linear dependence measures using a similar approach.
			Note that, although the purpose of this work is explicitly for linear dependence measures between time-series data, the modified $\Lambda$-test can be easily extended to more general likelihood tests for ratios of generalised variances under spatial autocorrelation~\cite{clifford1989assessing}, which is also known to affect the sampling properties of statistics.
			We begin by deriving tests for the (conditional) mutual information between both univariate and multivariate normally distributed processes.

	\section{Modified tests for mutual information}
	
	\label{sec:exact-tests-mutual-information}

	In this section, we obtain hypothesis tests for the mutual information between multiple time series.
	We first present the hypothesis tests explicitly for conditional mutual information for bivariate time series and, by using the chain rule, obtain the null distribution for the mutual information between multivariate time series.

	\subsection{Two time series}

	The conditional mutual information for linear Gaussian processes (Eq.~\eqref{eq:inf-cmi}) is equivalent to the statistic in Eq.~\eqref{eq:stat-l1}. Therefore, estimates of conditional mutual information under the null hypothesis, $\mathcal{H}_0 : \mathcal{I}_{X;Y \mid \mb{W}} = 0$, are $\Lambda^*$-distributed:
	\begin{align} \label{eq:dist-mi-exact}
		\exp{ \left( - 2\, \hat{\mathcal{I}}_{x;y \mid \mb{w}} \right)} &= 1-r_{xy\cdot \mb{w}}^2 \nonumber \\
                                                                        &\sim \Lambda^*(n_{xy\mid \mb{w}}).
	\end{align}
	Of course, due to the relationship between the $F$- and $\Lambda$-distribution (noted in Eq.~\eqref{eq:stat-l1}), we can construct an equivalent modified $F$-test for conditional mutual information:
	\begin{equation} \label{eq:dist-mi-bart-f}
		n_{xy\mid \mb{w}}\,[\exp{ ( 2\, \hat{\mathcal{I}}_{x;y\mid \mb{w}} )}-1] \sim F(1,n_{xy \mid \mb{w}}).
	\end{equation}
	The null distributions we provide above explicitly account for autocorrelation via the effective degrees of freedom $n_{xy\mid \mb{w}}$ and also reduce to the $F$-distribution for information-theoretic quantities when observations of the analysed time series are independent (cf.\ Eq.~\eqref{eq:dist-cmi-f} by letting $\hat{\eta}(x,y) = T$).
	Further, when $x$ and $y$ are serially uncorrelated, and in the limit $T\to \infty$, Eq.~\eqref{eq:dist-mi-bart-f} becomes equivalent to the $\chi^2$ null distribution for mutual information (see the discussion in \app{f-test}).
	Thus, the null distribution we present in Eq.~\eqref{eq:dist-mi-bart-f} is a generalisation of both the standard $F$-test (which is applicable only for i.i.d.\ variables) as well as the asymptotic distribution (which is applicable only for infinite data).
	
	Although they are special cases of the modified $\Lambda$-test, there are important distinctions here from the $\chi^2$-tests~\eqref{eq:dist-cmi-asymp} and the standard $F$-tests~\eqref{eq:dist-cmi-f} for conditional mutual information. The first is that we now have an effective sample size $\hat{\eta}$ that changes depending on the autocorrelation function of the residuals $e_{x \mid \mb{w}}$ and $e_{y \mid \mb{w}}$. The second is that the degrees of freedom $n_{xy\mid \mb{w}}$ is further reduced by $c=\dim{(\mb{w}(t))}$, the dimension of the conditional time series $\mb{w}$, which appears in the finite-sample $F$-tests but not the asymptotic $\chi^2$-tests. Both of these differences introduce a significant bias in the estimation of linear dependence for many real-world applications, exemplified by the numerical simulations in \secRef{mi-simulations}.

	\subsection{Multiple time series}

	Mutual information $\mathcal{I}_{\mb{X};\mb{Y} \mid \mb{W}}$ can also be used to measure the dependence between multivariate processes $\mb{X}$ and $\mb{Y}$. Here, we apply the chain rule and the results from the previous section to obtain a partial correlation decomposition that can be used for constructing a null distribution in the presence of autocorrelation.


	The chain rule provides a decomposition of mutual information as a sum of conditional mutual information terms:
	\begin{equation} \label{eq:chain-rule}
		\hat{\mathcal{I}}_{\mb{x};\mb{y} \mid \mb{w}} = \sum_{g=1}^k \sum_{h=1}^l \hat{\mathcal{I}}^{\{gh\}}_{\mb{xy\mid w}}.
	\end{equation}
	That is, mutual information estimates $\hat{\mathcal{I}}_{\mb{x};\mb{y} \mid \mb{w}}$ between a $k$-variate process $\mb{x}$ and an $l$-variate process $\mb{y}$, in the context of the $c$-variate concomitant $\mb{w}$, can be computed by summing over conditional mutual information terms~\cite{cover2012elements}.
	Each conditional mutual information term may be expressed as
	\begin{equation}
	    \hat{\mathcal{I}}^{\{gh\}}_{\mb{xy\mid w}} = \hat{\mathcal{I}}_{x_g ; y_h \vert \mb{v}^{\{gh\}}_{\mb{xy \mid w}}},
	\end{equation}
	where the conditional for the $(g,h)$-term is given by
	\begin{equation} \label{eq:cond}
		\mb{v}^{\{gh\}}_{\mb{xy\mid w}} = \begin{bmatrix} \mb{x}_{1:g-1} \\ \mb{y}_{1:h-1} \\ \mb{w} \end{bmatrix},
	\end{equation}
	with $\mb{x}_{1:g} = \begin{bmatrix}x_{1}', \ldots, x_{g}' \end{bmatrix}'$ a $g\times T$ matrix when $0 < g \leq k$, and the empty set, $\mb{x}_{1:g} = \emptyset$, when $g=0$.
	Using this notation, we have an equivalent expression for mutual information as
	\begin{align} \label{eq:inf-mvmi}
		\exp{ \left( - 2\, \hat{\mathcal{I}}_{\mb{x};\mb{y} \mid \mb{w}} \right) } &= \frac{\lvert s_{\mb{xy} \mid \mb{w}} \rvert}{\lvert s_{\mb{x} \mid \mb{w}} \rvert \lvert s_{\mb{y} \mid \mb{w}} \rvert} \nonumber \\
		&= \prod_{g,h} \exp{ \left(-2 \, \hat{\mathcal{I}}^{\{gh\}}_{\mb{xy\mid w}} \right) } \nonumber \\
		&= \prod_{g,h} \left( 1- r^2_{x_g  y_h \cdot \mb{v}^{\{gh\}}_{\mb{xy \mid w}}} \right).
	\end{align}
	Although this equation has a similar form to the well-known canonical correlation decomposition \cite{brillinger2004some,pham2008exact}, the correlations are over different variables.
	More importantly for our purposes, and assuming independence of these partial correlations (see below), its null distribution can thus be obtained from the $\Lambda^*$-distribution:
	\begin{equation} \label{eq:dist-mvmi-exact}
	    \exp{ \left( - 2\, \hat{\mathcal{I}}_{\mb{x};\mb{y} \mid \mb{w}} \right) } \sim \Lambda^*(\mb{n_{\mb{x}\mb{y}\mid \mb{w}}}),
	\end{equation}
	with parameter vector
	\begin{equation}
    	\mb{n}_{\mb{xy}\mid \mb{w}} = \left( n_{\mb{x}\mb{y}\mid \mb{w}}^{\{11\}}, \ldots, n_{\mb{x}\mb{y}\mid \mb{w}}^{\{kl\}} \right)'.
	\end{equation}
	The remaining challenge is to compute the effective degree of freedom, $n_{\mb{x}\mb{y} \mid \mb{w}}^{\{gh\}}$, for each independent partial correlation in Eq.~\eqref{eq:inf-mvmi}.
    Recall that partial correlation can be computed from ordinary least squares. The residual vector for the $(g,h)$ term in Eq.~\eqref{eq:inf-mvmi} is
	\begin{gather}
		e_{\mb{x} \mid \mb{v}}^{\{gh\}} = x_g - \hat{x}_g\Big(\mb{v}_{\mb{xy\mid w}}^{\{gh\}}\Big) \\
		e_{\mb{y} \mid \mb{v}}^{\{gh\}} = y_h - \hat{y}_h\Big(\mb{v}_{\mb{xy\mid w}}^{\{gh\}}\Big),
	\end{gather}
	where the hat $\hat{x}_g(\cdot)$ again denotes the linear prediction of $x_g$ from the input argument. Then, the degrees of freedom used in computing the $(g,h)$ sample partial correlation is:
	\begin{equation}
		n_{\mb{x}\mb{y}\mid \mb{w}}^{\{gh\}} = \hat{\eta} \! \left(e_{\mb{x} \mid \mb{v}}^{\{gh\}},e_{\mb{y} \mid \mb{v}}^{\{gh\}}\right) - \dim \!\left( \mb{v}_{\mb{x}\mb{y}\mid \mb{w}}^{\{gh\}}(t) \right) - 2.
	\end{equation}
	Throughout this analysis, we ordered the summations first over the dimensions of $\mb{y}$, and then over the dimensions of $\mb{x}$.
	In practice, the order of these operations are arbitrary and was initially imposed solely for clarity in the chain-rule formula.

	It should be noted that the modified $\Lambda$-test assumes both that the residuals are completely independent and that the estimated degrees of freedom is approximately correct.
	If, instead, the residuals become slightly correlated due to statistical errors in the regression, the $\Lambda^*$-distribution should be generated by sampling dependent beta- or $F$-distributed variables.
	Further, the effective degree of freedom is a first-order approximation, which may introduce biases in the hypothesis tests.
	Although our numerical simulations in \secRef{numerical-simulations} show no such biases, we discuss the potential solutions in Appendix~\ref{sec:apx-mv-bartletts}.

	\section{Modified tests for Granger causality}
	
	\label{sec:exact-tests-granger-causality}

	Recall that Granger causality can be expressed as a conditional mutual information (see Eq.~\eqref{eq:mvgc}). As such, we can leverage results from the previous section to introduce its null distribution. Many other information-theoretic and likelihood ratio-based measures could be similarly decomposed (from Wilks' criterion or conditional mutual information) in order to derive their finite-sample hypothesis tests.

	\subsection{Two time series}

	We shall first express the Granger causality for bivariate processes as a sum of conditional mutual information terms via the chain rule. Let upper indices (without parenthesis) denote a backshifted variable, e.g., $\mb{X}^{j}(t) = \mb{X}(t-j)$ denotes the variable $\mb{X}(t)$ lagged by $j$ time indices.
	Then, by applying the chain rule~\eqref{eq:chain-rule} to Granger causality~\eqref{eq:mvgc}, we can compute it as a sum of conditional mutual information estimates:
	\begin{equation} \label{eq:inf-gc}
		\hat{\mathcal{F}}_{y \to x \mid \mb{w}}(p,q) = 2\, \sum_{j=1}^q \hat{\mathcal{I}}_{x ;y^{j} \mid \mb{v}^{\{j\}}_{y\to x \mid \mb{w}}},
	\end{equation}
	with the $j$th conditional as the matrix
	\begin{equation}
		\mb{v}^{\{j\}}_{y\to x} = \begin{bmatrix} \mb{x}^{(p)} \\ \mb{y}^{(j-1)} \\ \mb{w} \end{bmatrix},
	\end{equation}
	and the limiting case giving $\mb{y}^{(0)} = \emptyset$, i.e., the empty set.
	Again, following Eq.~\eqref{eq:dist-mvmi-exact} we conclude that under the null hypothesis $\mathcal{H}_0 : \mathcal{F}_{Y\to X}(p,q) = 0$, Granger causality estimates are distributed as follows:
	\begin{equation} \label{eq:dist-gc-exact}
		\exp{\left(-\hat{\mathcal{F}}_{y\to x\mid \mb{w}}(p,q)\right)} \sim \Lambda^*(\mb{n}_{y\to x \mid \mb{w}}),
	\end{equation}
	where
	\begin{equation}
	    \mb{n}_{y\to x \mid \mb{w}} = \left(n_{y\to x \mid \mb{w}}^{\{1\}}, \ldots, n_{y\to x \mid \mb{w}}^{\{q\}} \right)'.
	\end{equation}

	Now, we can use the same approach from \secRef{exact-tests-mutual-information} to obtain the effective degrees of freedom used in computing Granger causality estimates. First, the residuals for the $j$th partial correlation in Eq.~\eqref{eq:inf-gc} are:
	\begin{align}
		e_{x\mid \mb{v}_{y\to x \mid \mb{w}}}^{\{j\}} &= x - \hat{x}\Big(\mb{v}_{y\to x\mid \mb{w}}^{\{j\}} \Big) \\
		e_{y\mid \mb{v}_{y\to x\mid \mb{w}}}^{\{j\}} &= y^{j} - \hat{y}^{j}\Big(\mb{v}_{y\to x\mid \mb{w}}^{\{j\}} \Big). 
	\end{align}
	Thus, the number of degrees of freedom is different for each term, with the $j$th number computed as:
	\begin{equation}
		n_{y\to x\mid \mb{w}}^{\{j\}} = \hat{\eta} \! \left(e_{x\mid \mb{v}_{y\to x\mid \mb{w}}}^{\{j\}} , e_{y\mid \mb{v}_{y\to x\mid \mb{w}}}^{\{j\}}\right) - \dim{\! \left( \mb{v}_{y\to x\mid \mb{w}}^{\{j\}}(t)\right)} -2.
	\end{equation}
	Unlike the standard $F$-test for Granger causality (Eq.~\eqref{eq:dist-gc-f}), the modified $\Lambda$-test takes into account the effective number of degrees of freedom induced by autocorrelation in both the predictee and predictor processes, with the two approaches overlapping only when there is no autocorrelation in the residuals. This indicates that, with limited data, the $F$-test can only be used for assessing the significance of Granger causality estimates from \emph{independent observations} ($y$) to univariate autocorrelated time series ($x$).

	\subsection{Multiple time series}

	Finally, we present hypothesis tests for the most complex linear dependence measure in the paper: the Granger causality from an $l$-variate predictor process $\mb{Y}$ to a $k$-variate predictee process $\mb{X}$, in the context of the $c$-variate concomitant process $\mb{W}$.
	By virtue of the chain rule~\eqref{eq:chain-rule}, this general expression of Granger causality~\eqref{eq:mvgc} decomposes into three nested sums:
	\begin{equation} \label{eq:inf-mvgc}
		\hat{\mathcal{F}}_{\mb{x} \to \mb{y} \mid \mb{w}}(p,q) = 2\,\sum_{g=1}^{k} \sum_{h=1}^{l} \sum_{j=1}^{q} \hat{\mathcal{I}}_{x_g y^{j}_h \mid         \mb{v}_{\mb{y} \to \mb{x}\mid \mb{w}}^{\{ghj\}} },
	\end{equation}
	where the conditional for the $(g,h,j)$ mutual information term is:
	\begin{equation}
		\mb{v}_{\mb{y} \to \mb{x}\mid \mb{w}}^{\{ghj\}} = \begin{bmatrix} \mb{x}_{1:g}^{(p)} \\[2mm] \mb{y}_{1:h-1}^{(q)} \\[2mm] \mb{y}_{h}^{(j-1)} \\ \mb{w} \end{bmatrix}.
	\end{equation}
	That is, the $(g,h,j)$ term is the conditional Granger causality for dimension $g$ of $\mb{X}$
	and the predictor observation $j$ steps back of the $h$th subprocess of $\mb{Y}$. This is conditioned on all dimensions and predictor observations below $g$, $h$, and $j$, as well as on $\mb{W}$.
	Again, following Eq.~\eqref{eq:dist-mvmi-exact} the distribution of Granger causality estimates, under the null hypothesis $\mathcal{H}_0:\mathcal{F}_{\mb{X}\to \mb{Y}}(p,q)=0$, is given as:
	\begin{equation} \label{eq:dist-mvgc-exact}
		\exp{ \left( \hat{\mathcal{F}}_{\mb{y}\to \mb{x}\mid \mb{w}}(p,q) \right) } \sim \Lambda^*( \mb{n}_{\mb{y}\to \mb{x}\mid \mb{w}} ),
	\end{equation}
	where
	\begin{equation} \nonumber
		\mb{n}_{\mb{y}\to \mb{x}\mid \mb{w}} = \left( n_{\mb{y}\to \mb{x} \mid \mb{w}}^{\{111\}}, \ldots, n_{\mb{y}\to \mb{x} \mid \mb{w}}^{\{klq\}} \right)'.
	\end{equation}
	The effective degrees of freedom are, again, computed from the residuals, where the residual processes used in computing the ($g,h,j$) partial correlation are:
	\begin{align}
		e_{\mb{x} \mid \mb{v}_{\mb{y} \to \mb{x} \mid \mb{w} }}^{\{ghj\}} &= x_{g} - \hat{x}_{g} \! \left(\mb{v}_{\mb{y} \to \mb{x} \mid \mb{w}}^{\{ghj\}} \right) \\
		e_{\mb{y} \mid \mb{v}_{\mb{y} \to \mb{x} \mid \mb{w}}}^{\{ghj\}} &= y_{h}^{j} - \hat{y}_{h}^{j} \! \left(\mb{v}_{\mb{y} \to \mb{x} \mid \mb{w}}^{\{ghj\}} \right).
	\end{align}
	The number of degrees of freedom for each term in the chained sum can then be computed from these residuals:
	\begin{equation}
		n_{\mb{y}\to \mb{x} \mid \mb{w}}^{\{ghj\}} = \hat{\eta}\!\left( e_{\mb{x} \mid \mb{v}_{\mb{y} \to \mb{x}\mid \mb{w}}}^{\{ghj\}},e_{\mb{y} \mid \mb{v}_{\mb{y} \to \mb{x}\mid \mb{w}}}^{\{ghj\}}\right) - \dim{\!\left(\mb{v}_{\mb{y} \to \mb{x}\mid \mb{w}}^{\{ghj\}}(t)\right)} -2.
	\end{equation}
	We note that, although the same $p$ and $q$ are used for each of the subprocesses of $\mb{x}$ and $\mb{y}$ in our presentation, our decomposition facilitates setting an individual history length for each term in the chained sum.
	The only difference would be to infer the optimal history length ($p$ and $q$) for each residual vector in the chain.


	\section{Numerical validation}
	\label{sec:numerical-simulations}

	We perform numerical simulations in order to validate the modified $\Lambda$-test and characterize the effect of autocorrelation on the (unmodified) $F$- and $\chi^2$-tests.
	The simulations, detailed in \app{numerical-simulations}, involve generating observations from two first-order independent \ac{AR} processes and iteratively filtering the output signal such that the autocorrelation is increased for both time series.

	Following Barnett and Seth~\cite{barnett2011behaviour}, we illustrate the finite-sample effects by generating relatively short stationary time series with $T=2^9=512$ observations from the stochastic processes to obtain our dataset.
	We begin by sampling first-order \ac{AR} processes to obtain our time-series data, $\mb{x}$, $\mb{y}$, and $\mb{w}$.
	Then, to illustrate the effect of higher autocorrelation on the \ac{FPR} of both methods, we digitally filter each time series along the time dimension with two types of low-pass causal filters: a \ac{FIR} linear-phase least-squares filter and an \ac{IIR} Butterworth filter.
	The filter order is variable, with higher filter orders generally increasing the autocorrelation of the time series.
	For each experiment, we perform 1\,000 trials and, using the statistical hypothesis-testing procedure described in \app{drawing-inferences}, we consider the \ac{FPR} to be the proportion of $p$-values that are significant at the nominal level (typically 5\% in this paper) in comparison to the relevant null distribution.
	
	These experiments allow us to study how each test behaves under increasing levels of autocorrelation of both $\mb{x}$ and $\mb{y}$, whilst ensuring that the null hypothesis (of no dependence) is not violated. 
    Rather than using a filter, it would of course be possible to increase the autocorrelation by selecting the ARMA parameters, $\mb{\Phi}(u)$ and $\mb{\Theta}(u)$, for each lag $u$.
    However, the formulations are equivalent: \ac{AR} processes are all-pole \ac{IIR} filters; \ac{MA} processes are \ac{FIR} filters; and ARMA processes are \ac{IIR} filters with both poles and zeros.
    Thus, although these are identical formulations, we opt for digitally filtering processes to increase their autocorrelation as this is a commmon preprocessing step performed by practitioners to remove artefacts from time series (even differencing the signal is a type of filter).
    Moreover, as previously discussed, filtering the signals has been shown in the past to bias various dependence measures such as Granger causality~\cite{barnett2011behaviour}.
    Until this work, however, it has not been suggested that this bias is a function of autocorrelation nor has a valid hypothesis test been proposed based on the autocorrelation function.

\subsection{Mutual information tests for bivariate time series}
\label{sec:mi-simulations}

	\begin{figure}[t]
	\centering
		\subfloat[]{
			\includegraphics[width=\columnwidth]{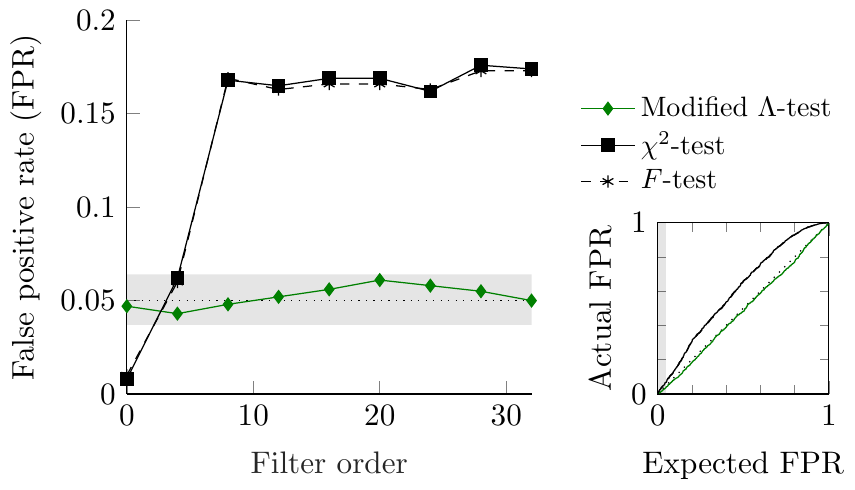}
			\label{fig:mi-tests-fir}
		}
		\\
		\subfloat[]{
			\includegraphics[width=\columnwidth]{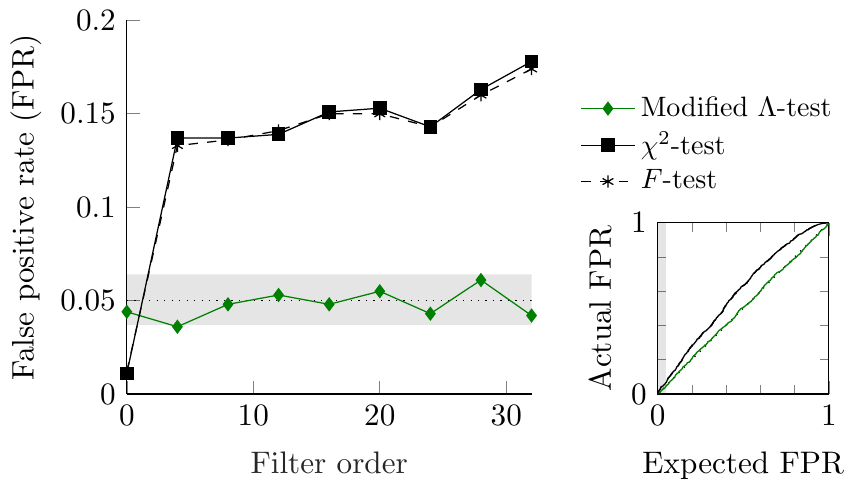}
			\label{fig:mi-tests-iir}
		}
	\caption{
	    Modified tests correctly assess the significance of the mutual information estimated between two univariate time series under FIR (\ref{fig:mi-tests-fir}) and IIR (\ref{fig:mi-tests-iir}) filtering.
		The mutual information was measured (using Eq.~\eqref{eq:inf-cmi}) between independent univariate time series (generated by Eq.~\eqref{eq:mv-model}) after filtering, and tested using the $\chi^2$-test (Eq.~\eqref{eq:dist-cmi-asymp}), $F$-test (Eq.~\eqref{eq:dist-cmi-f}), and the modified $\Lambda$-test (Eq.~\eqref{eq:dist-mi-exact}). The plots show the effect of increasing the filter order on the \ac{FPR} for both an (a) FIR filter and (b) IIR filter.
		The shaded regions on the right indicate $\alpha=0.05$, whilst the shaded regions on the left show the 95\% confidence interval for the FPR (as defined in \app{drawing-inferences}).
		The subplots on the right capture the \ac{FPR} for all potential significance levels $\alpha$ (``Expected FPR'') with an $8$th order filtered signal. An ideal distribution is where the \ac{FPR} equals $\alpha$ and thus sits perfectly on the diagonal, as per our tests.
	}
	\label{fig:mi-tests}
\end{figure}

\begin{figure}[t!]
		\centering
		\subfloat[]{
			\includegraphics[width=\columnwidth]{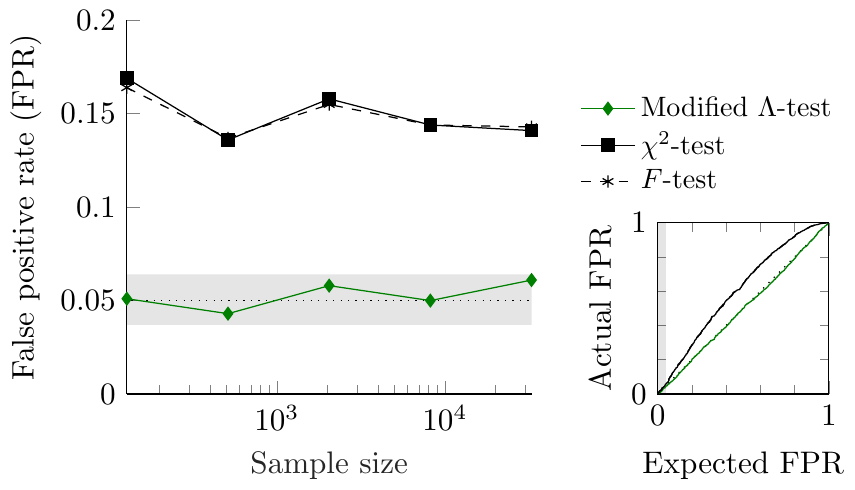}
			\label{subfig:mi-tests-large-sample-fir}
		}
		\\
		\subfloat[]{
			\includegraphics[width=\columnwidth]{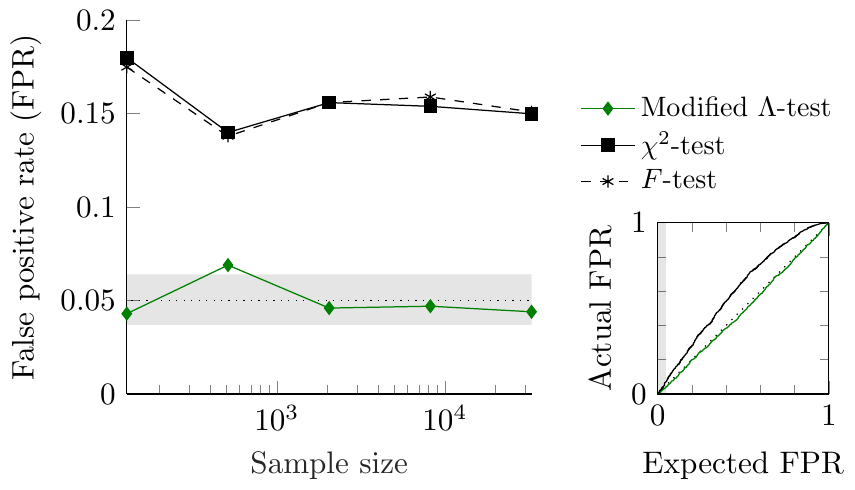}
			\label{subfig:mi-tests-large-sample-iir}
		}
	\caption{
		Increasing the sample size does not mediate the effect of autocorrelation on the $\chi^2$- and $F$-tests for mutual information.
		We perform the same tests as \fig{mi-tests}, except with an exponentially increasing sample size and a fixed 8th-order FIR (\ref{subfig:mi-tests-large-sample-fir}) and  IIR (\ref{subfig:mi-tests-large-sample-iir}) filter.
		The subplots on the right show the FPR for $T=2^{11}$ samples.
		}
	\label{fig:mi-tests-large-sample}
	\end{figure}

First, we use this approach to evaluate the performance of the hypothesis tests on assessing the significance of mutual information estimates between two independent (but serially correlated) time series.
Our results are shown in \fig{mi-tests}, where the ``$F$-tests'' are from the finite-sample distribution (Eq.~\eqref{eq:dist-cmi-f}), ``$\chi^2$-tests'' refer to the asymptotic \ac{LR} distributions (Eq.~\eqref{eq:dist-cmi-asymp}), and the ``Modified $\Lambda$-tests'' refer to our hypothesis tests that account for autocorrelation (Eq.~\eqref{eq:dist-mi-exact}). As the plots illustrate, both the $F$-tests and the $\chi^2$-tests overestimate the measures for higher filter orders (and therefore higher \ac{AR} orders), yielding over $15\%$ of false positives at the nominal significance of $\alpha = 0.05$---approximately three times the \ac{FPR} expected from the test. The figures on the right illustrate the significance level $\alpha$ against the \ac{FPR} for an $8$th order filter. From these figures, we can see that the \ac{FPR} for the $\chi^2$-tests is higher than nominal for all significance levels $\alpha \in (0,1)$. In comparison, the modified $\Lambda$-test procedure yields the expected \ac{FPR} for all filter orders.

A filter order of zero in \fig{mi-tests} refers to generating the time-series data with the first-order \ac{AR} model~\eqref{eq:mv-model} without any digital filtering.
In this case, the $\chi^2$- and $F$-tests yielded less than the nominal 5\% \ac{FPR}.
This occurs when the number of effective samples becomes greater than the original sample size $\eta(x,y) > T$.
Referring to Bartlett's formula~\eqref{eq:bartletts}, this is due to the product of negative autocorrelation exhibited by the $Y$ process (induced by $\Phi_Y = -0.8$, indicating an effect of undersampling) and the positive autocorrelation exhibited by the $X$ process (induced by $\Phi_X = 0.3$).
Counter-intuitively, this would imply that the observations are anti-correlated in time.
Such conservative results for the $\chi^2$- and $F$-tests are likely to induce lower statistical power (i.e., a lower \ac{TPR}) in scenarios when the effective sample size is greater than the original sample size.
To verify this, we performed 1\,000 trials where there was a small dependence of $X$ on $Y$ (see \app{numerical-simulations}).
The \ac{TPR} was 0.049 (SE of 0.0068) for the $\chi^2$-test and 0.1570 (SE of 0.0115) for the modified $\Lambda$-test.
Thus, the power of our hypothesis test is three times greater than the $\chi^2$-test in this scenario.

In \fig{mi-tests-large-sample} we show that increasing the sample size does not mediate the effect of autocorrelation on the $\chi^2$- and $F$-tests.
This is due to the fact that the effective degree of freedom is always a fraction of the degree of freedom.
Thus, regardless of the sample size, both the asymptotic ($\chi^2$) and finite ($F$) tests are invalid, unless modified to account for effective sample size.

\subsection{Conditional mutual information tests for bivariate time series}

\label{sec:cmi-simulations}

\begin{figure}[t]
	\centering
		\subfloat[]{
			\includegraphics[width=\columnwidth]{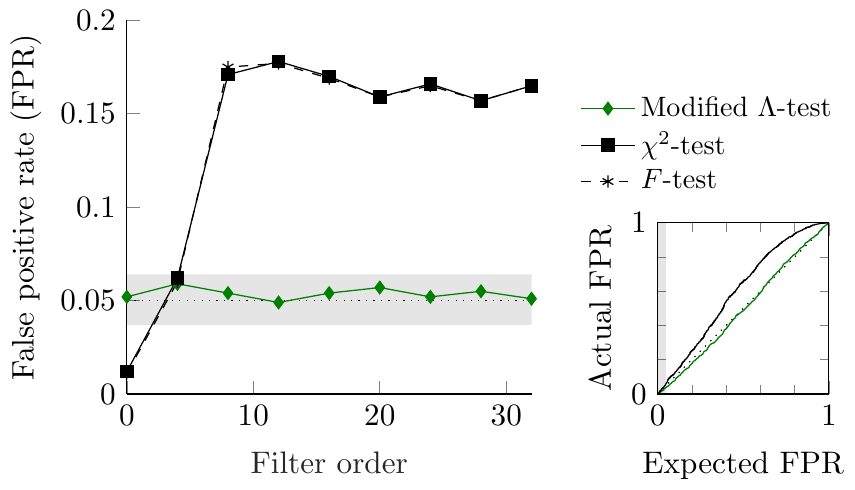}
			\label{subfig:cmi-fir-z1}
		}
		\\
		\subfloat[]{
			\includegraphics[width=\columnwidth]{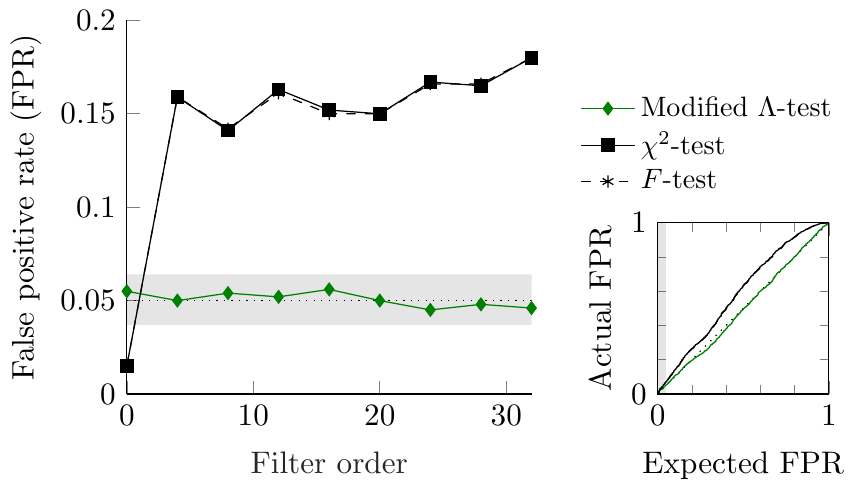}
			\label{subfig:cmi-iir-z1}
		}
	\caption{
	    Modified tests correctly assess the significance of the conditional mutual information estimated between two univariate time series, conditioned on a third univariate time series; each time series underwent FIR (\ref{subfig:cmi-fir-z1}) or IIR (\ref{subfig:cmi-iir-z1}) filtering.
		Conditional mutual information was measured (using Eq.~\eqref{eq:inf-cmi}) between two univariate time series, given a third univariate process (generated by Eq.~\eqref{eq:mv-model} with $k=l=c=1$), after filtering, and tested for significance using the $\chi^2$-, $F$-, and modified $\Lambda$-tests.
		The subplots on the right show the \ac{FPR} for each significance level $\alpha$ when the signal is filtered with an $8$th order filter.}
	\label{fig:cmi-tests}
\end{figure}

\begin{figure}
        \centering
		\subfloat[FIR filter with $c$-variate conditional.]{
			\includegraphics[width=\columnwidth]{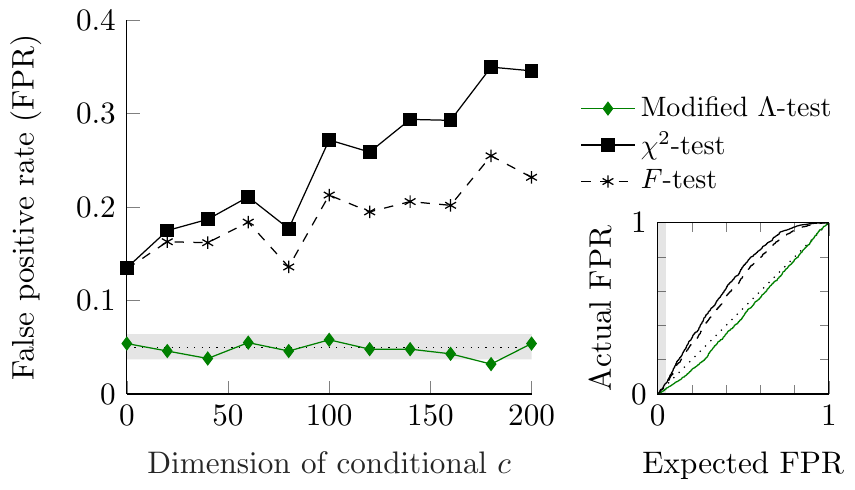}
			\label{subfig:cmi-fir}
		}
		\\
		\subfloat[IIR filter with $c$-variate conditional.]{
			\includegraphics[width=\columnwidth]{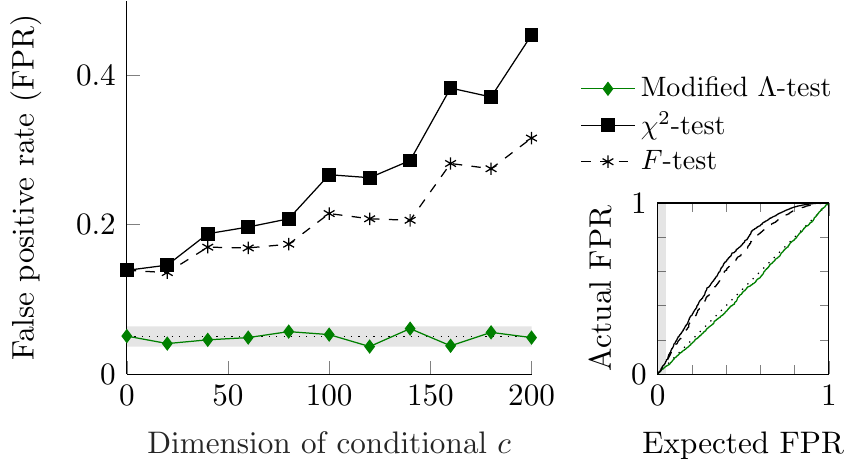}
			\label{subfig:cmi-iir}
		}
	\caption{
	    The \ac{FPR} of asymptotic tests increase with the dimension of the $c$-variate conditional process.
	    The plots are the same as per \fig{cmi-tests}, except as a function of $c$ with a fixed $8$th order FIR (\ref{subfig:cmi-fir}) and IIR filter (\ref{subfig:cmi-iir}).
		The subplots on the right show the \ac{FPR} for each significance level $\alpha$ when $c = 100$.
	}
	\label{fig:cmi-tests-increasing-c}
\end{figure}

We now extend the previous results by evaluating the effect of conditioning mutual information between $x$ and $y$ on an independent, tertiary process $\mb{w}$.
The \acp{FPR} for the $\chi^2$-tests, $F$-tests, and the modified $\Lambda$-tests from these experiments are presented in \fig{cmi-tests}, and exhibit similar characteristics to those of the mutual information tests in \fig{mi-tests}.
Increasing the filter order generally increases the \ac{FPR} for both unmodified tests, yet the modified $\Lambda$-test remains unbiased, maintaining the expected \ac{FPR} of 5\%.

As discussed in \secRef{exact-tests-mutual-information}, an important distinction between the null distributions for mutual information (Eq.~\eqref{eq:dist-mi-exact}) is that the effective degree of freedom in the $\Lambda^*$-distribution not only includes the effective sample size but also the dimension of the conditional $c$.
To show the severity of the asymptotic approximation, we generate the $x$ and $y$ processes and filter the signal with an $8$th order FIR and IIR filter the same as before; however, we increase the number of independent processes $c = \dim{(\mb{W}(t))}$ in the multivariate conditional.
The result is shown in \fig{cmi-tests-increasing-c}, where the \ac{FPR} of the $F$- and $\chi^2$-tests increases somewhat linearly with the dimension, however the modified $\Lambda$-test remains unbiased.
As evidenced by the improvemenet of the unmodified $F$-test over the $\chi^2$-test, this experiment demonstrates that even when the autocorrelation function is the same, the dimension of the conditional must also be included in the hypothesis tests.

\subsection{Mutual information tests for multivariate time series}
\label{sec:mvmi-simulations}

\begin{figure}[t!]
	\centering
	\subfloat[]{
		\includegraphics[width=\columnwidth]{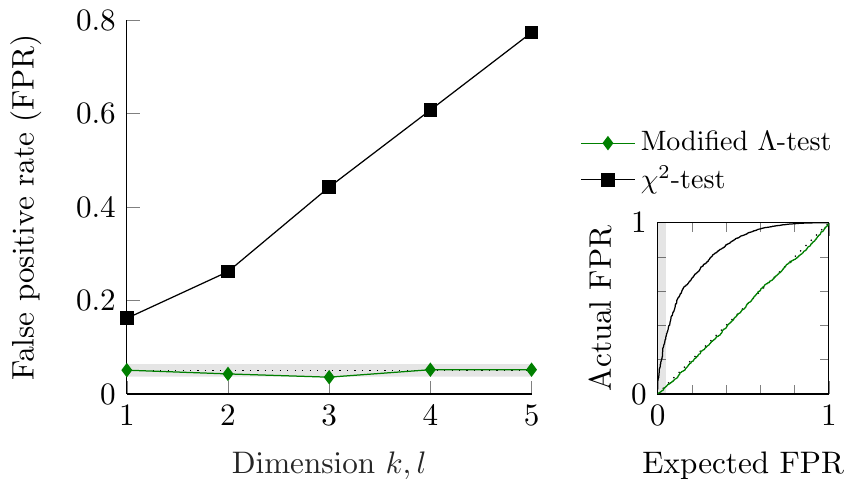}
		\label{subfig:mv-mi-cond-tests-fir}
	}
	\\
	\subfloat[]{
		\includegraphics[width=\columnwidth]{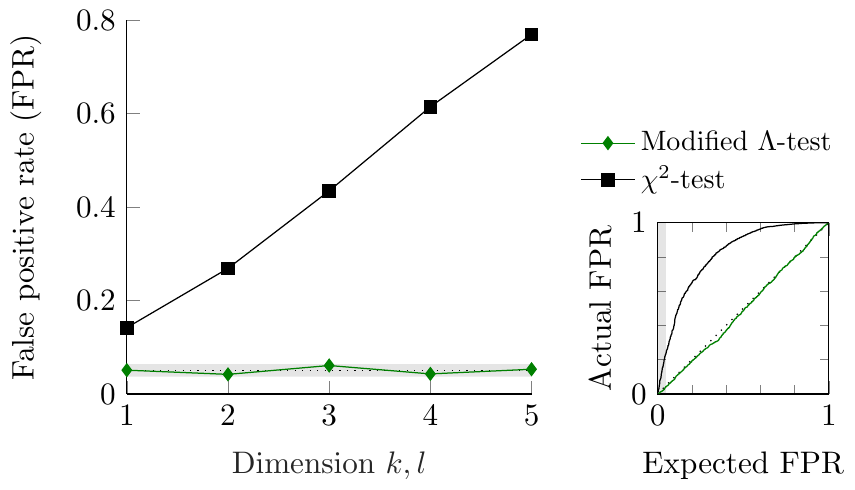}
		\label{subfig:mv-mi-cond-tests-iir}
	}
	\caption{
	    Modified $\Lambda$-tests correctly assess the significance of the mutual information estimated between two multivariate time series.
		The mutual information was measured (using Eq.~\eqref{eq:inf-mvmi}) between the multivariate time series (generated by Eq.~\eqref{eq:mv-model} with an increasing dimension $k$ and $l$ of $\mb{X}$ and $\mb{Y}$), passed through 8th-order FIR (\ref{subfig:mv-mi-cond-tests-fir}) and IIR (\ref{subfig:mv-mi-cond-tests-iir}) filters, and tested for significance using the $\chi^2$-test (Eq.~\eqref{eq:dist-cmi-asymp}) and the modified $\Lambda$-test (Eq.~\eqref{eq:dist-mvmi-exact}). The subplots on the right show the \ac{FPR} for each significance level $\alpha$ when $k = l = 3$.}
	\label{fig:mv-mi-cond-tests}
\end{figure}

In this section, we present results for the hypothesis tests of mutual information between multivariate ($m>2$) time series.
The multivariate time series are partitioned into two independent sets of processes, $\mb{X}$ and $\mb{Y}$, one with dimension $k$ and one with dimension $l$.
For each experiment, we let $k=l$ and use the state equations in Eq.~\eqref{eq:mv-model} to simulate $m=k+l$ independent AR processes for $m=\{2,4,\ldots,10\}$. These signals are then filtered along the temporal dimension using $8$th order \ac{FIR} and \ac{IIR} filters. This signal generation process ensures that there is no correlation between signals within the same subprocess, i.e., $\rho_{ij} = 0$ for all $i,j \in [1,m]$. The results are shown in \fig{mv-mi-cond-tests}, where increasing the dimension approximately linearly increases the \ac{FPR} of the original \ac{LR} test to over 70\% for both filters (continuing to increase for larger $k$ and $l$), yet the modified $\Lambda$-test remains unbiased.

In the tests above, no correlations between subprocesses were included (e.g., $X_i(t)$ with $X_j(t-u)$ for $j\neq i$), however an internal cross-correlation between any of these subprocesses may further decrease the size and power of unmodified tests. Our experiments of Granger causality in the following sections naturally incorporate examples with correlated subprocesses in the mutual information calculation.

\subsection{Granger causality tests for bivariate time series}

\label{sec:gc-simulations}

\begin{figure}[t!]
	\centering
	\subfloat[]{
		\includegraphics[width=\columnwidth]{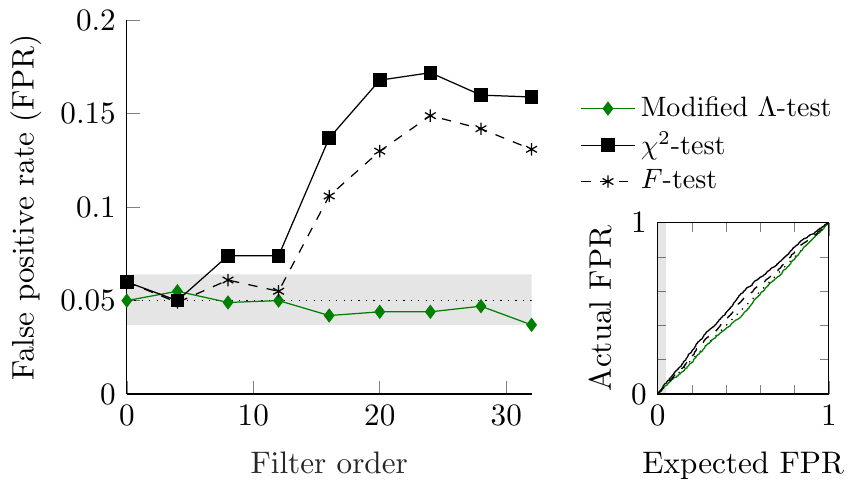}
		\label{subfig:te-fir}
	}
	\\
	\subfloat[]{
		\includegraphics[width=\columnwidth]{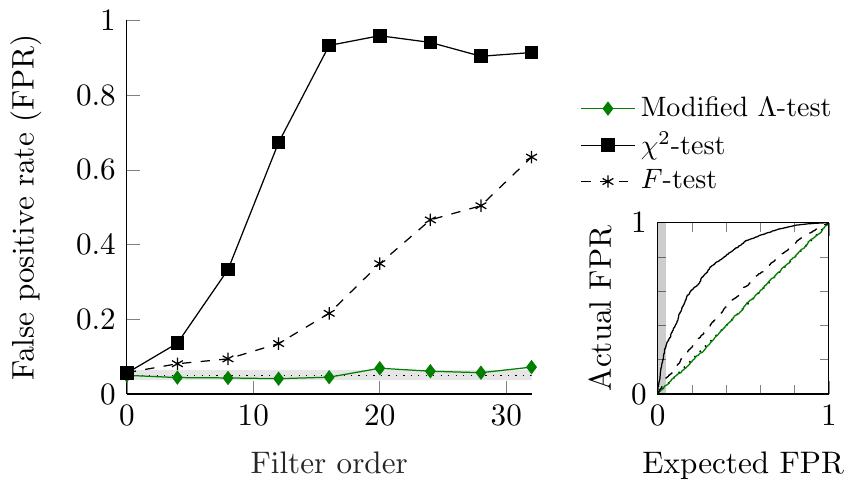}
		\label{subfig:te-iir}
	}
	\caption{
		Modified $\Lambda$-tests correctly assess the significance of the Granger causality estimated from one univariate time series to another. Granger causality, with history lengths $p$ and $q$ chosen via Burg's method, is estimated (using Eq.~\eqref{eq:inf-gc}) between univariate time series (generated by Eq.~\eqref{eq:mv-model}) after smoothing with an FIR (\ref{subfig:te-fir}) and an IIR (\ref{subfig:te-iir}) filter.  Estimates are then tested using the $\chi^2$-test (Eq.~\eqref{eq:dist-mvgc-asymp}), the $F$-test (Eq.~\eqref{eq:dist-gc-f}), and the modified $\Lambda$-test (Eq.~\eqref{eq:dist-gc-exact}). The subplots on the right show the \ac{FPR} for each significance level $\alpha$ with an 8th-order filter.}
	\label{fig:gc-tests}
\end{figure}

\begin{figure}
    \centering
	\subfloat[]{
		\includegraphics[width=\columnwidth]{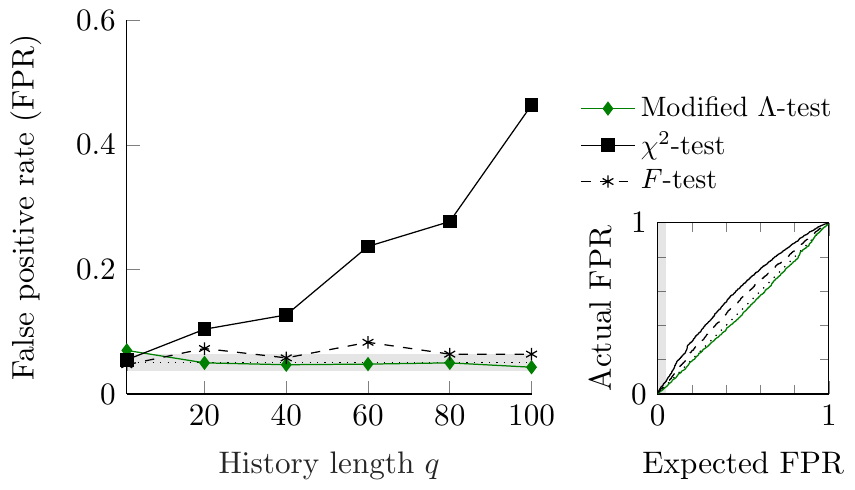}
		\label{subfig:te-fir-h}
	}
	\\
	\subfloat[]{
		\includegraphics[width=\columnwidth]{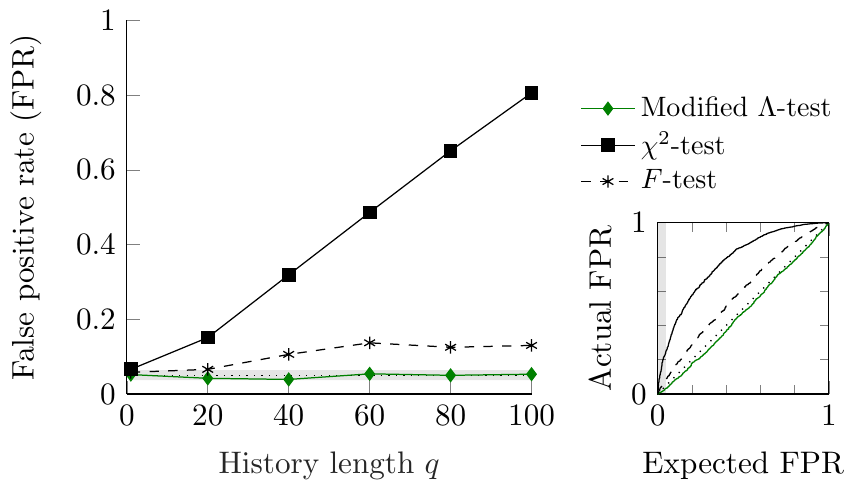}
		\label{subfig:te-iir-h}
	}
	\caption{
	    The FPR of the $F$- and $\chi^2$-tests for Granger causality increases with an increasing dependence on the past for time series generated by an 8th-order FIR (\ref{subfig:te-fir-h}) and IIR (\ref{subfig:te-iir-h}) filter.
	    We illustrate this by varying the history length of the predictor process, $q$, from one to $200$.
	    The subplots on the right show the \ac{FPR} for each significance level $\alpha$ when $q = 20$ (for Fig.~\ref{subfig:te-fir-h}) and $q=40$ (for Fig.~\ref{subfig:te-iir-h}) to approximately match the orders chosen via Burg's method in Fig.~\ref{fig:gc-tests}.
	}
	\label{fig:gc-tests-q}
\end{figure}


This section examines the performance of each hypothesis test on estimates of Granger causality using the same (univariate) simulations from \secRef{mi-simulations}. That is, the bivariate AR model (Eq.~\eqref{eq:mv-model} with $k=l=1$ and no conditional $c=0$) is simulated to generate $T=512$ observations of the $X$ and $Y$ processes, which are then passed through \ac{FIR} and \ac{IIR} filters. Referring to \fig{gc-tests}, we perform this with each filter order and each filter type (\ac{FIR} and \ac{IIR}). After generating these sample paths, the \ac{AR} order of the predictee, $p$, and the predictor, $q$, were inferred from Burg's method~\cite{de1996yule}. We then compute Granger causality (via Eq.~\eqref{eq:inf-gc}) and use this estimate to obtain $p$-values from the CDFs of the $F$-distribution (Eq.~\eqref{eq:dist-gc-f}), the $\chi^2$-distribution (Eq.~\eqref{eq:dist-mvgc-asymp}), and the $\Lambda^*$-distribution (Eq.~\eqref{eq:dist-gc-exact}). This is performed 1\,000 times in order to obtain a \ac{FPR} of each approach.
Whilst our experiment here is equivalent to a conditional mutual information, unlike the experiments in previous sections the autocorrelation within the time series naturally induces a cross-correlation amongst the variables within the predictor $\mb{Y}^{(q)}(t)$ and conditional $\mb{X}^{(p)}(t)$ processes.
The results shown here illustrate that increasing the autocorrelation length via filtering increases the \ac{FPR} of Granger causality under the $F$- and $\chi^2$-tests, particularly when using an IIR filter.
In contrast, the \ac{FPR} of Granger causality using the modified $\Lambda$-tests remains mostly unbiased.
It should be noted that, although within the confidence bounds, the FPR of the modified $\Lambda$-tests appear to be not exact for high-order filters; sources of error regarding this potential bias are discussed in Appendix~\ref{sec:apx-mv-bartletts}.

\begin{figure}[t!]
		\centering
		\subfloat[]{
			\includegraphics[width=\columnwidth]{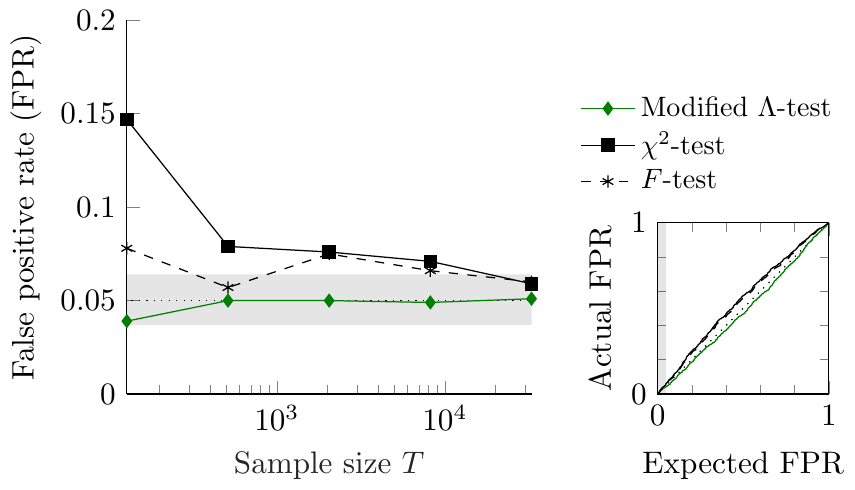}
			\label{subfig:gc-tests-large-sample-fir}
		}
		\\
		\subfloat[]{
			\includegraphics[width=\columnwidth]{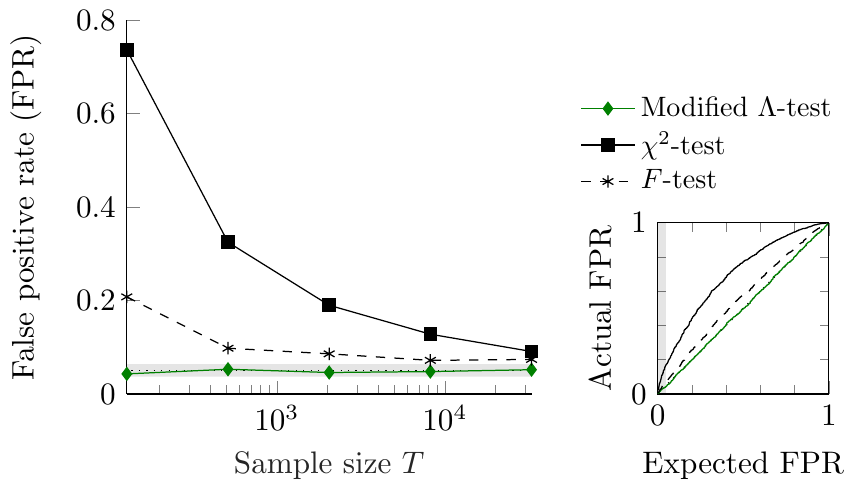}
			\label{subfig:gc-tests-large-sample-iir}
		}
	\caption{
		The $F$- and $\chi^2$-tests converge to the modified $\Lambda$-test for Granger causality with large sample sizes.
		This experiment is the same as \fig{gc-tests}, except with an exponentially increasing sample size and a fixed 8th-order FIR (\ref{subfig:gc-tests-large-sample-fir}) and IIR (\ref{subfig:gc-tests-large-sample-iir}) filter.}
	\label{fig:gc-tests-large-sample}
	\end{figure}
	
		\begin{figure}[t!]
		\centering
		\subfloat[]{
			\includegraphics[width=\columnwidth]{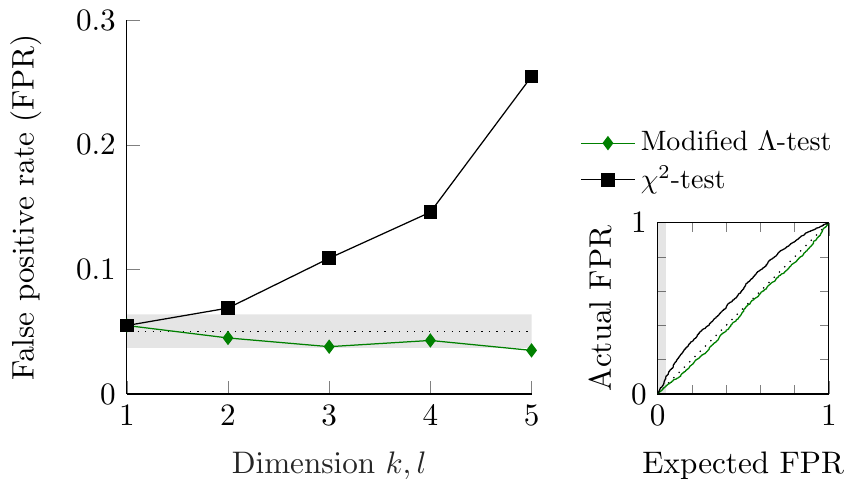}
			\label{subfig:mvgc-fir}
		}
		\\
		\subfloat[]{
			\includegraphics[width=\columnwidth]{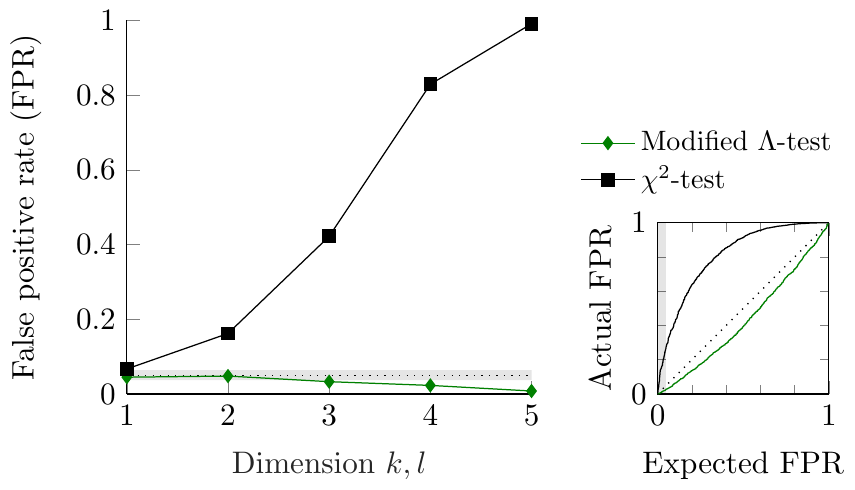}
			\label{subfig:mvgc-iir}
		}
	\caption{
	    Modified $\Lambda$-tests correctly assess the significance of the Granger causality estimated from one multivariate time series to another for increasing dimension $k,l$.
		Granger causality, with each predictee history length chosen optimally, is estimated (using Eq.~\eqref{eq:inf-mvgc}) between multivariate time series (generated by Eq.~\eqref{eq:mv-model}) after filtering via 8th-order FIR (\ref{subfig:mvgc-fir}) and IIR (\ref{subfig:mvgc-iir}) filters.
		Estimates are then tested using the $\chi^2$-tests (Eq.~\eqref{eq:dist-mvgc-asymp}) and our modified $\Lambda$-test (Eq.~\eqref{eq:dist-mvgc-exact}). The subplots on the right show the \ac{FPR} for each significance level $\alpha$ when $k = l = 3$.}
	\label{fig:mvgc}
	\end{figure}

The model order for our experiments above was chosen using Burg's method, however, the are numerous approaches to inferring the ``optimal'' AR order as outlined in the introduction, all of which can result in vastly different model orders.
To ensure our results are not a consequence of poor model identification, in \fig{gc-tests-q} we illustrate the FPRs for increasing the predictor history length $q$ from one to 200 (in increments of 20), whilst holding the autocorrelation length constant with an 8th-order filter.
This effectively introduces more terms in Eq.~\eqref{eq:inf-gc}, causing a larger divergence between the $F$-, $\chi^2$- and modified $\Lambda$-tests.
As expected, the \ac{FPR} of the $\chi^2$- and $F$-tests linearly increases in this range, whereas the modified $\Lambda$-test remains consistent with the $5\%$ \ac{FPR}. This linear increase of the \ac{FPR} in $\chi^2$- and $F$-tests is somewhat counter-intuitive to the notion of Granger causality, where one may expect that accounting for more history would reduce spurious correlations. However, the opposite is true, simply due to a lack of correct finite-sample distributions (in the case of the unmodified tests).

In \fig{gc-tests-large-sample} we show the effect of increasing the sample size for tests on Granger causality estimates.
Here, we can see the $\chi^2$- and $F$-tests converging for sufficient sample sizes.
Unlike mutual information (from \fig{mi-tests-large-sample}), estimating Granger causality involves regressing the autocorrelation of the predictee first, with the variance of these residuals reducing as the sample size grows.
Thus, the effective sample size asymptotically approaches the sample size, however, the precise rate of this convergence depends the autocorrelation function and may change for every pair of time series.
Even for this simple example, we see that on the order of 100 000 samples are required for convergence, which is not realistic in many empirical scenarios.

\subsection{Granger causality tests for multivariate time series}

\label{sec:mvgc-simulations}

Finally, we can evaluate the effect of increasing the dimensionality of both processes on Granger causality inference.
In these experiments, we vary the dimension of the processes $\mb{X}$ and $\mb{Y}$ from one to five. Recall from Eq.~\eqref{eq:inf-mvgc} that the number of terms involved in computing Granger causality (and its null distribution) is the product $klq$, of the dimensionality ($k,l$) and the history length of the predictor ($q$).
Due to the relatively short time series length of $T=512$ samples and high autocorrelation and dimensionality, allowing an arbitrary predictor history length of $q$ results in the effective sample size approaching zero for the modified $\Lambda$-tests. Thus, for these experiments we fix the history length of the predictor to $q=1$.

The results are shown in \fig{mvgc}, where the $\chi^2$-tests inflate the \ac{FPR} close to 100\% for with higher dimensional processes. Although our corrected tests perform well for moderate dimensionality, when $k,l > 3$ with the IIR filter, the \ac{FPR} of our modified $\Lambda$-tests begin to have numerical issues. This is caused by the regression matrix not being well conditioned, i.e., the ratio of regressors to data points is too high. Nonetheless, we can see from the figure that our tests maintain a low \ac{FPR}, becoming more conservative when the regression is ill-posed. Moreover, a poorly conditioned regression can be easily tested for in practice. So, we conclude that with minimally sufficient observations, our tests maintain the desired \ac{FPR} even for the most general case of multivariate Granger causality and, when the sample size is simply too small for reliable inference, our approach flags this as an issue.

\section{Effect of prewhitening}
\label{sec:prewhitening}

\begin{figure}[t!]
		\centering
		\subfloat[]{
			\includegraphics[width=\columnwidth]{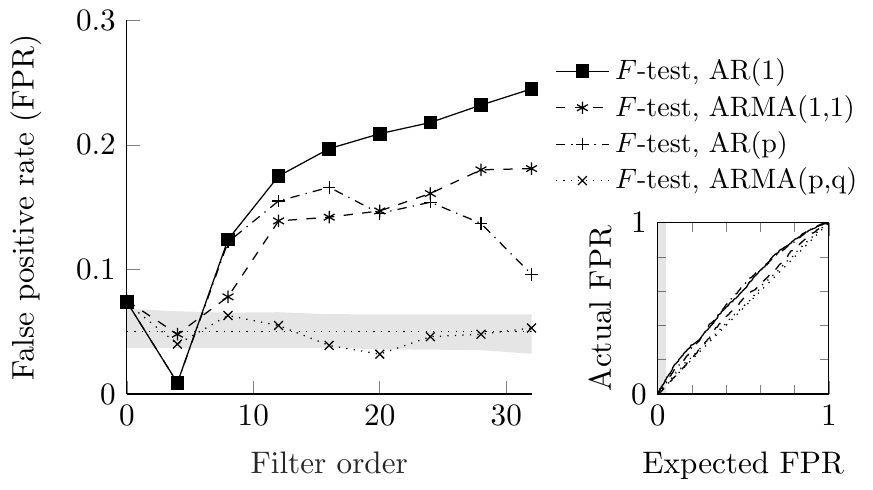}
			\label{subfig:mi-prewhitening-fir}
		}
		\\
		\subfloat[]{
			\includegraphics[width=\columnwidth]{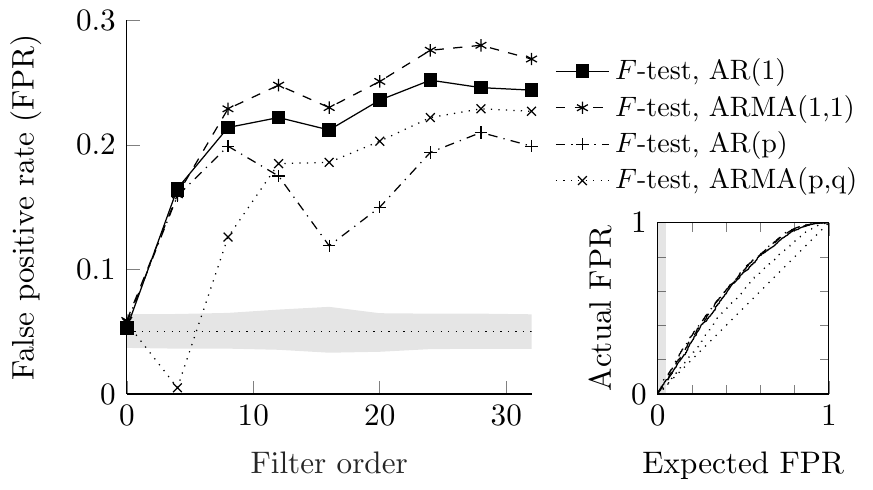}
			\label{subfig:mi-prewhitening-iir}
		}
	\caption{
		Prewhitening the time series does not mediate the bias in $F$-tests for mutual information estimates after FIR (\ref{subfig:mi-prewhitening-fir}) or IIR (\ref{subfig:mi-prewhitening-iir}) filtering.
		The experiments are as per \fig{mi-tests}, with four prewhitening approaches used: the $\text{AR}(1)$, $\text{ARMA}(1,1)$, $\text{AR}(p)$, and $\text{ARMA}(p,q)$.
		For the $\text{AR}(p)$ and $\text{ARMA}(p,q)$ models, the optimal order is inferred from Burg's method and the BIC score.
		In many cases for the IIR-filtered time series, either the $\text{AR}(p)$ or the $\text{ARMA}(p,q)$ failed to learn a stable model, and so these trials were removed, inducing non-uniform confidence intervals (reflected in the shaded region).}
	\label{fig:mi-prewhitening}
\end{figure}

\begin{figure}[t!]
		\centering
		\subfloat[]{
			\includegraphics[width=\columnwidth]{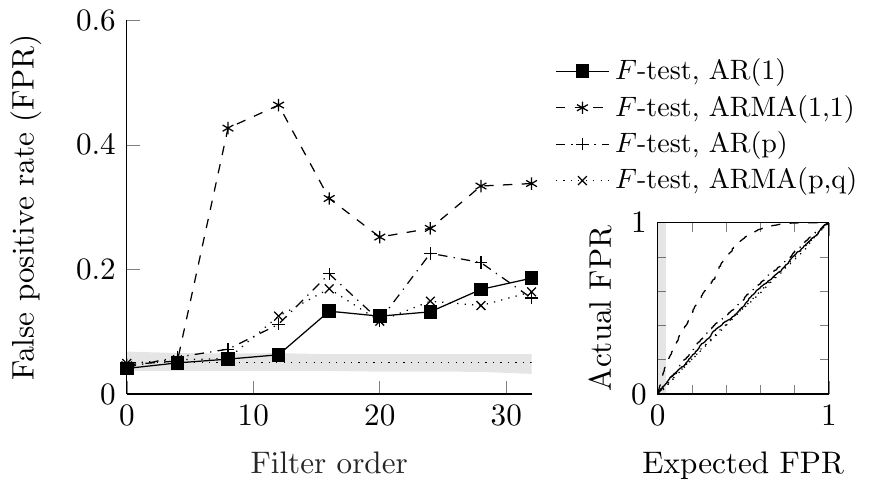}
			\label{subfig:gc-prewhitening-fir}
		}
		\\
		\subfloat[]{
			\includegraphics[width=\columnwidth]{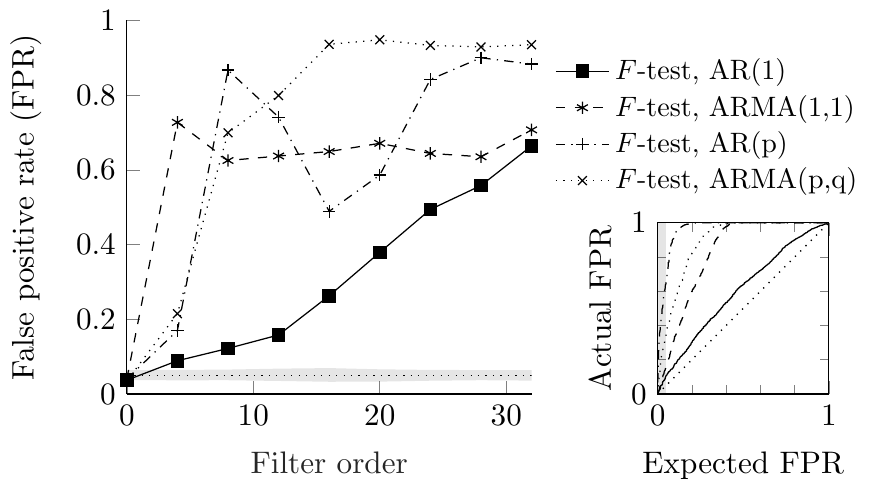}
			\label{subfig:gc-prewhitening-iir}
		}
	\caption{
		Prewhitening the time series does not mediate the bias in $F$-tests for Granger causality estimates after FIR (\ref{subfig:gc-prewhitening-fir}) or IIR (\ref{subfig:gc-prewhitening-iir}) filtering.
		The experiments are as per \fig{gc-tests}, with four prewhitening approaches used (as per \fig{mi-prewhitening}): the $\text{AR}(1)$, $\text{ARMA}(1,1)$, $\text{AR}(p)$, and $\text{ARMA}(p,q)$.}
	\label{fig:gc-prewhitening}
\end{figure}

The rationale for applying prewhitening is to remove the autocorrelation in one time series such that the variance of computed statistics becomes equivalent to serially independent observations~\cite{cryer2008time}.
That is, instead of modifying the hypothesis tests, prewhitening modifies the input time series and thus the statistics themselves.
Prewhitening is typically attempted by first inferring a model of one time series ($\mb{x}$), and transforming the process $\mb{x}$ to a residual process $\tilde{\mb{x}}$ through a filter constructed from its model.
The same filter (with parameters inferred from $\mb{x}$) is then applied to the other time series $\mb{y}$ to create $\tilde{\mb{y}}$.
Specifically, assuming any arbitrary $\textrm{ARMA}(p,q)$ model for time series $\mb{x}$, prewhitening involves learning the parameter vectors $\hat{\mb{\Phi}}$ and $\hat{\mb{\Theta}}$ from $\mb{x}$, and then filtering the raw signals through the following equations:
\begin{align}
    \tilde{\mb{x}}(t) &= \mb{x}(t) - \sum_{u=1}^p \hat{\mb{\Phi}}(u) \mb{x}(t-u) - \sum_{u=1}^q \hat{\mb{\Theta}}(u) \tilde{\mb{x}}(t-u), \nonumber \\
    \tilde{\mb{y}}(t) &= \mb{y}(t) - \sum_{u=1}^p \hat{\mb{\Phi}}(u) \mb{y}(t-u) - \sum_{u=1}^q \hat{\mb{\Theta}}(u) \tilde{\mb{y}}(t-u). \nonumber
\end{align}
Using the same linear transformation (filter) for both time series renders their correlation theoretically invariant~\cite{cryer2008time}, however, the assumption is that $\tilde{\mb{x}}$ is now serially independent, and so the variance of sample correlations (for instance) converge to $\hat{\sigma}_r(x,y) = 1/T$.

A significant challenge in prewhitening signals is in selecting an appropriate model of the autocorrelation function.
Of course, for the same arguments as presented in \secRef{preliminaries}, after mean removal and differencing, the most general model for covariance-stationary time series are ARMA models.
However, inference procedures to learn both the order and parameters of ARMA models are computationally expensive.
As such, many authors (and textbooks~\cite{cryer2008time}) propose an $\textrm{AR}(p)$ model would suffice to render the residuals, $\tilde{\mb{x}}$, independent, presuming that autocorrelations decay rapidly for stationary time series.
Since this is the same assumption underlying Granger causality (regarding the residuals on the target after fitting an $\textrm{AR}(p)$ model), our results from \secRef{gc-simulations} suggest that statistics computed from signals prewhitened in this way will remain biased.
In fMRI research, the most popular packages that are used for preprocessing time-series data (AFNI, SFL, and SPM) are similarly insufficient for handling autocorrelation due to their simplistic models~\cite{olszowy2019accurate}.
Specifically, the package AFNI uses an ARMA(1,1) model learned from each voxel, whereas FSL uses Tukey tapering to smooth the data (see Appendix~\ref{sec:apx-mv-bartletts}), and SPM uses one global AR(1) model for all processes.
Thus, each package assumes a fixed ARMA model can describe any arbitrary-order process.
This is clearly insufficient, and if the wrong model is used, the residuals $\tilde{\mb{x}}$ remain dependent, resulting in an unknown variance of statistical estimates.
This is evidenced by consistently high \acp{FPR} in empirical studies~\cite{olszowy2019accurate}.

For completeness of this paper, however, we implement a number of prewhitening schemes in order to illustrate that such an approach is insufficient for assessing linear dependence between typical time series.
Our experiments---on the same synthetic time series used in \secRef{numerical-simulations}---show the effect of prewhitening univariate signals on the unmodified $F$-test for a number of different models: the $\textrm{AR}(1)$ and $\textrm{ARMA}(1,1)$; as well as the $\textrm{AR}(p)$ and $\textrm{ARMA}(p,q)$ with the optimal model orders ($p$ and $q$) learned from data.
For the $\textrm{AR}(p)$ model inference, we allow for $p \in [1,200]$ (where the sample size is $T=512$), with the model order selected (and parameters inferred) using Burg's method~\cite{de1996yule}.
Given the difficulty of learning higher-order $\textrm{ARMA}(p,q)$ models, however, we restrict our search space, iterating through each potential $p,q \in \{1,\ldots,5\}$ and selecting the model with the lowest BIC score.

\begin{figure*}[t!]
	\centering
	\includegraphics[width=\textwidth]{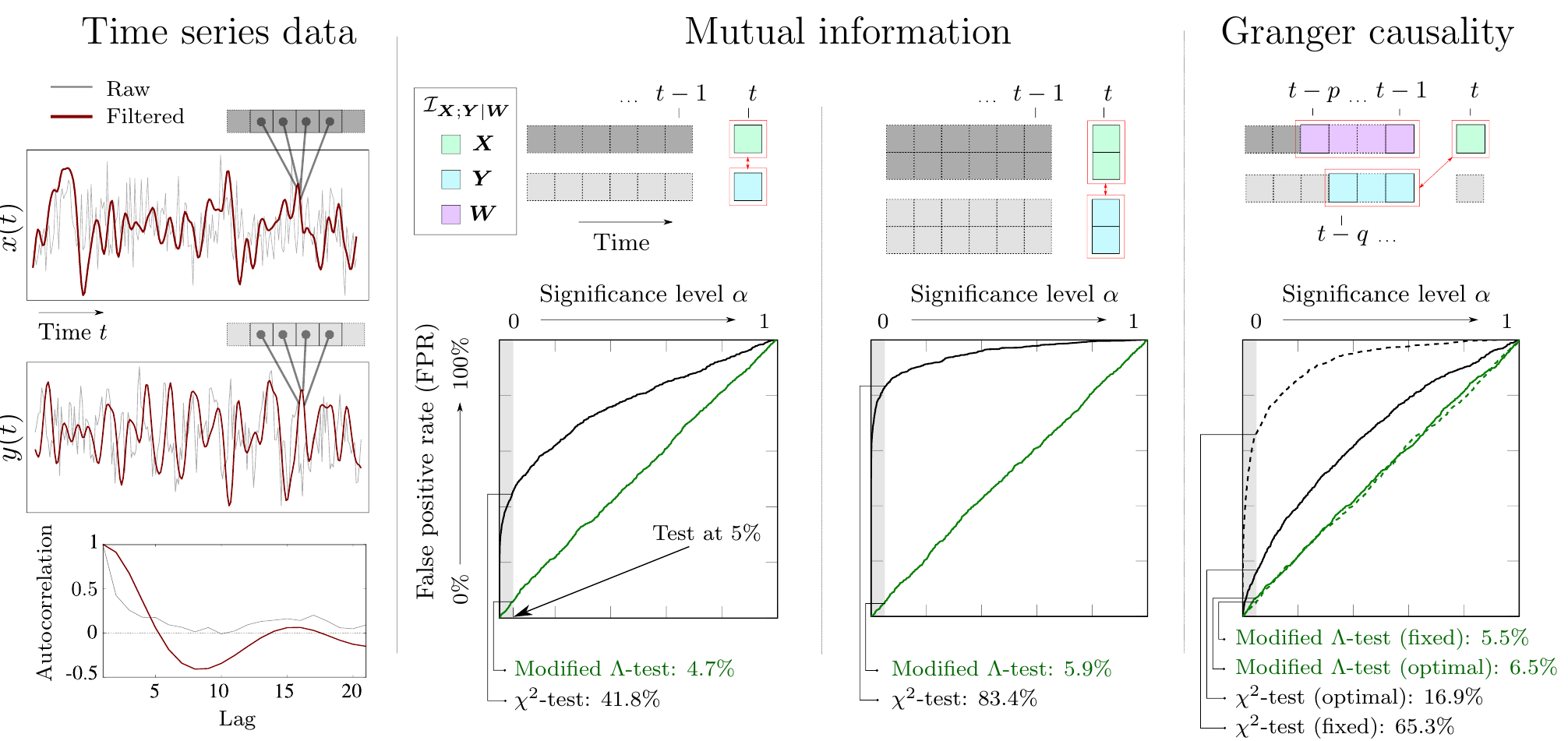}
	\caption{
		Application to brain-imaging data, demonstrating our correction for the otherwise dramatic inflation of false-positive rates (FPRs) of classical hypothesis tests of dependence estimates in the presence of autocorrelation. We perform 1\,000 experiments with both the $\chi^2$-tests as well as the modified $\Lambda$-tests for inferring the significance of two dependence measures: mutual information(between two univariate and two bivariate processes), and Granger causality. For each of these experiments, we randomly selected two uncorrelated fMRI time series, $\mb{X}$ and $\mb{Y}$, from the Human Connectome Project~\cite{van2012human} (see \app{fmri} for more details). A sample window of these univariate processes are shown in the left panel (grey lines), with the common preprocessing technique of band-pass digital filtering later applied to the signals (red lines). At the bottom of the left panel, we plot the sample autocorrelation function from both the raw and the filtered signal, illustrating the higher autocorrelation length (and lower effective sample size) induced by digital filtering. The panels on the right show the \ac{FPR} of each test applied to the filtered signals as a function of the significance level $\alpha$. For ideal hypothesis testing, we expect a line along the diagonal (i.e., the \ac{FPR} equals the significance level). The $\chi^2$-tests illustrate an increased \ac{FPR} for all measures, as seen by both the plots and the \ac{FPR} at 5\% significance level. In contrast, our tests remain consistent with the expected \ac{FPR}.}
	\label{fig:main}
\end{figure*}

Our results for the performance of the $F$-test on mutual information estimates after prewhitening are shown in \fig{mi-prewhitening}.
Contrasting these results to \fig{mi-tests}, the only benefit of prewhitening appears to be for the FIR-filtered time series when an $\text{ARMA}(p,q)$ model is used.
In almost all other scenarios, the \acp{FPR} are either equivalent to, or worse than, the original tests.
Figure~\ref{fig:gc-prewhitening} illustrates the effect of prewhitening on Granger causality $F$-tests (compare to \fig{gc-tests} without prewhitening).
Again, prewhitening appears to serve no benefit to tests for Granger causality, even for the relatively advanced $\text{ARMA}(p,q)$ model.
Concerningly, for IIR-filterd signals, the \ac{FPR} increases to over $60\%$ for all ARMA models with no scenario where the prewhitening approach results in an \ac{FPR} within the expected range.
In Appendix~\ref{sec:apx-prewhitening} we demonstrate similar results when using BIC and AICc scoring functions (as an alternative to Burg's method) to infer $\text{AR}(p)$ models for prewhitening.

It is possible that, with an ideal model, the $F$- or $\chi^2$-tests used for a prewhitened time series may be comparable (in size properties or FPR) to the modified $\Lambda$-test.
However, even restricting our search space to an $\textrm{ARMA}(5,5)$ model resulted in an approximately 5\,000-fold increase in computational time over the modified $\Lambda$-test~\footnote{Using the MATLAB \texttt{estimate} function in order to infer the parametres for each $p$ and $q$; see our open-source toolkit for details.} and, moreover, remained biased in most scenarios.
We conclude that, regardless of the model selected, the additional burden of prewhitening over using a modified hypothesis test is unjustified and that the outcome of unmodified hypothesis tests (with or without prewhitening) conveys inconsistent information about the underlying dependence structure between time series.

\section{Case study: human connectome project dataset}
\label{sec:fmri}

Studies in computational neuroscience often leverage statistical analysis in order to formulate and test biologically plausible models. An important application in this field is the study of the human brain through fMRI, which is abundant with short, autocorrelated time series that have been studied using the measures of interest here~\cite{barnett2011behaviour,davey2013filtering,afyouni2019effective,florin2010effect,seth2010matlab}. As such, it is an archetypal real-world application to illustrate the issues of autocorrelation for time-series analysis.

In fMRI research, the blood-oxygen level-dependent (BOLD) data is translated into a slowly varying (and thus highly autocorrelated) multivariate time series that traces the haemodynamic response of different locales (voxels) in the brain.
Digital filtering is then commonly used as a preprocessing step to reduce line noise, nonstationarity and other artefacts in neuroimaging data.
This induces an (either finite or infinite) impulse response that can increase autocorrelation, even if the original signals were not serially correlated.
To characterise the \ac{FPR} of linear dependence measures between empirical time series, we use completely independent time series from a widely accessed brain-imaging dataset known as the the Human Connectome Project (HCP) resting state fMRI (rsfMRI) dataset~\cite{van2012human} (see \app{fmri} for more detail; time series are selected from different random regions of interest from different random subjects to ensure independence, and then digitally filtered). This process is shown in \fig{main}, where a sample window of the raw and filtered data from two independent time series appears on the left panel.


First, we illustrate the effect of autocorrelation on mutual information by using a $\chi^2$-test, with significance level 5\%, computed with $T=800$ samples of each independent time series.
We begin by estimating mutual information $\hat{\mathcal{I}}_{x;y}$ between two unrelated time series $x$ and $y$, sampled from the HCP dataset.
We find that the test results are significantly biased, yielding an \ac{FPR} of 41.8\% (a 9-fold increase of the expected rate). After prewhitening with an $\textrm{AR}(p)$ filter, this increased to over 88\% (not shown in the figure).
The modified $\Lambda$-test demonstrated an \ac{FPR} of 4.7\%, which is well within the acceptable confidence interval.
Next, we perform the same tests but with the mutual information between two sets of bivariate time series $\mb{x}$ and $\mb{y}$ ($k=l=2$); this yields an 83\% \ac{FPR} (a 16-fold increase).
When we correctly test these same measurements with the modified $\Lambda$-test (Eq.~\eqref{eq:dist-mvmi-exact}), we find a \ac{FPR} of 5.2\%, matching the desired level within the confidence bounds.

For Granger causality, we perform two different tests: scenario (i) with the model orders, $p$ and $q$, inferred for each trial, and scenario (ii) with a fixed $p=q=100$ for all trials. 
For unrelated signal pairs, the \ac{FPR} of Granger causality estimates using the $\chi^2$-test is 16.9\% (with optimal history length around 18), increasing further to 90.5\% after prewhitening with an $\textrm{AR}(p)$ filter.
If a longer embedding length of 100 is chosen, the FPR is 66\%---more than thirteen times the expected value---increasing to 83.3\% after prewhitening with an $\mathrm{AR}(p)$ model.
When we test these same measurements with the modified $\Lambda$-test~\eqref{eq:dist-gc-exact}, we find a \ac{FPR} of 5.5\% and 6.5\% for scenario (i) and (ii), respectively, completely removing the false-positive bias exhibited by the $\chi^2$-tests.

\section{Discussion}
\label{sec:discussion}

We have shown that the autocorrelation exhibited in covariance-stationary time-series data induces bias in the hypothesis tests of a broad class of linear dependence measures.
By framing different dependence measures in unified theoretical terms, we provide the first demonstration of how Bartlett's formula can be applied to derive unbiased hypothesis tests, termed modified $\Lambda$-tests, for mutual information (and, consequently, Wilks' criterion and Granger causality) for both univariate and multivariate time-series data.
These measures are used in a wide range of disciplines, modelling myriad important processes from anthropogenic climate change~\cite{zhang2011causality} to the brain dynamics of dementia patients~\cite{franciotti2013default}.
The continued use of flawed testing procedures in empirical sciences is problematic, making it imperative that the corrections reported here be incorporated into future studies.

The effect of temporal autocorrelation on linear dependence has long been investigated in statistics, however the majority of research has focused on simple linear correlations~\cite{bartlett1935some,bartlett1946theoretical,roy1989asymptotic,davey2013filtering,afyouni2019effective}, which are restricted to measuring symmetric bivariate dependence structures.
These studies, while representing important milestones in inferring the association between univariate time series, suffer from the inability to capture both multivariate and directed dynamical dependence.
Even though measures such as mutual information and Granger causality can model a much richer set of dependence structures in temporal processes, the notion that their sampling properties could be altered in the presence of autocorrelation has thus far been largely overlooked.
A major challenge of handling autocorrelation for more involved dependence measures was in extending Bartlett's formula to multivariate relationships.
Crucially, our approach facilitates not only independent multivariate relationships but also correlated multivariate processes.
Our results on the mutual information between multivariate time series used examples where the subprocesses were all independent.
When extending this approach to Granger causality for arbitrarily large history lengths and dimensionality, the subprocesses become inherently correlated (since one subprocess is a time-lagged version of another, and they involve significant autocorrelation).
This provides strong evidence that our approach is a generalisation of Bartlett's dependence studies that is able to handle a much richer class of multivariate dependency structures beyond those already presented in this paper.

Granger causality is the \emph{de facto} measure of directed dependence between stationary time series.
The typical approach to assessing the significance of (potentially multivariate) estimates has been via the finite-sample $F$-tests~\cite{barnett2014mvgc} or the asymptotic $\chi^2$-test~\cite{geweke1982measurement}.
In this study, we have shown that higher levels of autocorrelation (equivalently, a higher-order autoregression) in the signal inflates the variance of these statistical estimates, inducing significant bias for these traditional hypothesis tests.
This means that using these tests induces errors when the predictor process is serially correlated---precisely the situation that Granger causality was designed to address.
One might logically surmise that this issue could be mediated by accounting for a longer history of the process, i.e., conditioning on additional \ac{AR} variables.
Referring to \fig{gc-tests}, we have shown that this will result in the \ac{FPR} being even further inflated.
This is particularly concerning given that the autocorrelation function is one of the two properties that define a stationary process~\cite{brockwell1991time,reinsel2003elements,box2015time}---the other is its mean, which has no effect on scale-invariant dependence measures such as Granger causality, mutual information, and Pearson correlation.
Similar to the thinking behind Granger causality, another common approach to remove autocorrelation is prewhitening, which intends to induce a serially independent process (of residuals) for hypothesis testing.
Our experiments in \fig{mi-prewhitening} and \fig{gc-prewhitening} (as well as Appendix~\ref{sec:apx-prewhitening}) illustrated that many common approaches to prewhitening fail to control the \ac{FPR} and, indeed, can further reduce the size of the unmodified ($F$ and $\chi^2$) hypothesis tests.
We conclude that the unmodified hypothesis tests cannot be used to reliably infer the significance of Granger causality or mutual information estimates when applied to covariance-stationary time series and should be replaced with modified tests, such as the modified $\Lambda$-tests, particularly for limited time-series~data.

Prior work~\cite{barnett2011behaviour} had already established that digital filtering led to biased Granger causality estimates for shorter time-series, yet this effect was not understood nor able to be corrected until now. Due to the widespread use and influence of Granger causality across fields including neuroscience, ecology and economics, underlined by any examination of the literature (see \secRef{introduction}), this was a serious deficiency for directed inference of relationships in time-series analysis. Much like correlation coefficients, the issue was magnified in fields dealing with short, highly autocorrelated time-series, as demonstrated in \fig{main} for computational neuroscience using fMRI recordings.
Many extensions to Granger causality have been proposed to explicitly model the autocorrelation function (such as ARMA~\cite{boudjellaba1992testing} or state-space Granger causality~\cite{barnett2015granger}).
However these approaches are also known to exhibit significant false-positive biases~\cite{gutknecht2019sampling}, aligning with our results in \secRef{prewhitening} on the ineffectiveness of prewhitening with ARMA models.
In this paper, we showed that modifying the effective degrees of freedom of the null distribution suffices to eliminate the bias across all examined time series, without the additional burden of inferring complex models or prewhitening.
More advanced methods (such as state-space Granger causality) may retain other empirical advantages, but their hypothesis tests are likely to require incorporation of similar modifications based on effective sample sizes.
More broadly, our results strongly suggest that any hypothesis tests dealing with time-series analysis should be modified to account for autocorrelation, regardless of regressing or conditioning on \ac{AR} components.
The concerningly high \acp{FPR} exhibited in our experiments suggest that relationships established using previously tested Granger causality estimates should perhaps be revisted; particularly in fields that have high levels of autocorrelation and limited data.

Throughout this work, we have made the assumption that the time-series innovations are Gaussian and that all relationships are linear.
Thus, we have only discussed linear-Gaussian probability distributions for information-theoretic measures.
When instead applied to nonlinear time series, these probability distributions are often inferred using non-parametric density estimation techniques such as nearest neighbour or kernel methods~\cite{lizier2014jidt}.
Spurious estimates of the nearest neighbour counts have previously been observed for autocorrelated signals by Theiler~\cite{theiler1986spurious}, who provided a solution by excluding observations that are close in time. This is now a popular approach to effectively account for autocorrelation in density estimation for nonlinear time-series analysis.
In fact, in introducing transfer entropy---now understood as a model-free extension of Granger causality~\cite{barnett2009granger}---Schreiber explicitly recommended the use of a Theiler window (also known as serial- or dynamic-correlation exclusion) when kernel estimation methods are used~\cite{schreiber2000measuring}. 
The Theiler window approach has been demonstrated to control the \ac{FPR} for such estimators in practice~\footnote{See the Supporting Information of Novelli et al.~\cite{novelli2019large} for a similar experiment with transfer entropy on the same HCP rsfMRI dataset used in our \fig{main}},
yet remains a heuristic with no theoretical guarantees and, similar to Pearson correlation, is often neglected in practical estimation of transfer entropy~\footnote{Weber et al.~\cite{weber2017influence} report a contrasting finding of increased \ac{FPR} in transfer entropy inference due to filtering, however the use of a Theiler window is not specified}.
We hypothesize that the methods outlined in our work could be extended in future to provide a more rigorous approach to handling autocorrelated nonlinear time series through, e.g., nonlinear versions of Bartlett's formula~\cite{francq2009bartlett}, facilitating a broader class of information-theoretic measures.

Finally, the dependence structure discussed in this work is assumed to be in the time domain, whereas many empirical studies are concerned with other forms of dependence that could similarly bias hypothesis tests.
Future work will be required to consider handling such correlation structures in a similar fashion to that which we have presented, e.g., spectral models~\cite{granger1969investigating} or spatial autocorrelation~\cite{clifford1989assessing}.
Indeed, the formula for the effective sample size (Eq.~\eqref{eq:bartletts}) was developed for spatial autocorrelation and can thus be easily extended to handle spatiotemporal autocorrelation, allowing for even broader class of null distributions that can be considered with the modified $\Lambda$-test.

	\appendix

	\section{Statistical hypothesis tests}
	\label{apx:hypothesis-tests}
	
	The linear dependence measures discussed in this paper are positive real-valued random variables $\hat{\Lambda} \in \mathbb{R}_{>0}$ that can be expressed as the ratio of the generalised variance of two models. In general, we consider two nested models, the `restricted' model with $p_0$ parameters (under which the null hypothesis $\mathcal{H}_0$ is true) and the `unrestricted' model with $p_1$ parameters (under which the alternate hypothesis $\mathcal{H}_1$ is true). These models are referred to as nested since $p_0 < p_1$ and the restricted model parameter space is a subset of the unrestricted model space. The statistics are expressed in terms of the generalised sample variance of these models:
	\begin{equation} \label{eq:statistic}
	\hat{\Lambda} = \frac{|\mb{s}_0|}{|\mb{s}_1|},
	\end{equation}
	where $\mb{s}_i$ is the the residual sum-of-squares for model $i$ and $|\mb{s}_i|$ is the generalised sample variance.
	The generalised sample variance is, asymptotically, inversely proportional to the likelihood of each model, and so taking the log of the ratio of generalised sample variances~\eqref{eq:statistic} is equivalent to the \ac{LR} between two models (for a large enough number of samples)~\cite{reinsel2003elements,boudjellaba1992testing}.
	For this reason, statistics of the form in Eq.~\eqref{eq:statistic} appear in a number of linear dependence measures, such as mutual information (with Gaussian marginals) and Geweke's definition of Granger causality~\cite{geweke1982measurement}.
	
	\subsection{Asymptotic likelihood-ratio test}
	
	Wilks' theorem~\cite{wilks1938large} is the basis of the $\chi^2$-test, which states that a test statistic constructed from the \ac{LR} of nested models will asymptotically follow a $\chi^2$-distribution under the null hypothesis.
	Since all statistics used in this work fit this definition, a $\chi^2$-test can be used.
	That is, if the true model is the restricted model, then as $T \to \infty$, the statistic is chi-square distributed:
	\begin{equation} \label{eq:chi2-test}
	    T \, \log{( \hat{\Lambda} )} \overset{d}{\to} \chi^2(p_1-p_0).
	\end{equation}
	However, as we show throughout the main text, the $\chi^2$-test has a significant bias when applied to limited and autocorrelated time-series data, which results in a large number of false positives.
	
	It is important to note that the \ac{LR} test is but one of three classical procedures for hypothesis testing maximum likelihood estimates; the others are the Wald test and the Lagrange multiplier test~\cite{brockwell1991time,box2015time,reinsel2003elements}.
	The three tests overlap because the null distribution of each asymptotically follows the $\chi^2$-distribution. Thus, the same issues hold if one were to use any test on linear dependence measures unless autocorrelation is considered in the null distributions.
	
	\subsection{Finite-sample \texorpdfstring{$F$}{F}-test}
	\label{app:f-test}
	
	In regression analysis, the $F$-test is used to infer the significance of nested models of independent observations with limited data, i.e., the finite-sample null distribution. Using the same notation as above, we obtain a distribution for the comparing the nested models:
	\begin{equation} \label{eq:f-test}
	\frac{T-p_1}{p_1-p_0} \frac{S_0 - S_1}{S_1} \sim F(p_1 - p_0, T - p_1 ).
	\end{equation}
	$F$-statistics can be reformulated as nested ratios of sample variances through simply rearranging the LHS of~\eqref{eq:f-test}, i.e.,
	\begin{align} \label{eq:f-test-lr}
		\frac{T-p_1}{p_1-p_0} \left[ \hat{\Lambda} - 1 \right] \sim F(p_1 - p_0, T - p_1 ).
	\end{align}
	Thus, the $F$-statistic is a function of the \ac{LR} of two models and we can show its asymptotic distribution is chi-square. First, a Taylor expansion of the LHS of the $F$-statistic in Eq.~\eqref{eq:f-test-lr} gives $\log{(\hat{\Lambda})} \approx \hat{\Lambda} - 1$. Moreover, for a random variable $X \sim F(\nu_1, \nu_2)$, then $Y = \lim_{\nu_2 \to \infty} \nu_1 X$ has the chi-square distribution $\chi^2(\nu_1)$. Thus, by this asymptotic relationship between the $F$- and $\chi^2$-distributions, we have:
	\begin{align}
		\lim_{T \to \infty} (T-p_1) \log{(\hat{\Lambda})} &\overset{d}{\to} ( p_1 - p_0 ) \, F(p_1 - p_0, T ), \\
						T \, \log{(\hat{\Lambda})} &\sim \chi^2(p_1-p_0).
	\end{align}
	This result is discussed throughout the paper to explain the diverging behaviour between the two tests.
	
	\subsection{Surrogate-distribution tests}
	
	Another established approach to empirically generating a null distribution involves permuting, re-drawing or rotating the observations of one variable $x$ or $y$ and computing the relevant statistic for each surrogate dataset~\cite{lizier2014jidt,barnett2011behaviour,anderson2001permutation}.
	Naive approaches to permuting or re-drawing will completely destroy the autocorrelation profile of that variable, making this empirical distribution representative of serially independent observations and similar to the analytic $F$-distribution.
	Indeed, such empirical generation of the CDF via permutation testing was attempted (for Granger causality) by Barnett and Seth \cite{barnett2011behaviour}, and shown to incur the same inflated FPR issues as the $F$- and $\chi^2$-tests.
	Alternatively, constrained realisation approaches~\cite{schreiber1996improved} can be used to generate surrogate time-series data that exhibit certain properties (such as the same power spectra) and have recently been shown effective for handling autocorrelation in EEG data~\cite{schaworonkow2015power}.
	Nonetheless, bootstrap tests are computationally inefficient and known to exhibit size and power distortions, however typically less so than asymptotic tests~\cite{davidson1999size}.
	As such, we consider a comparison to these empirical approaches outside the scope of our paper.

\subsection{Drawing inferences}
\label{app:drawing-inferences}
	
Given an estimate $\hat{\Lambda}$ and null distribution (e.g., the $F$- or $\chi^2$-distributions), we use the same general hypothesis testing procedure for all measures. Here, we use the $\chi^2$-test for $\hat{\Lambda}$ as an example, however the same applies to all linear dependence measures, including the $\Lambda^*$-distributions, which are numerically generated.

First, we set an arbitrary significance level $\alpha$, which is (ideally) the probability of rejecting the null hypothesis even if it were true---this is set to 5\% in this paper unless stated otherwise. Then, the statistic $T \, \log{( \hat{\Lambda} )}$ is input to the quantile function of the $\chi^2$-distribution with $p_1-p_0$ degrees of freedom. The output (the complement of the $p$-value) is the probability of measuring that value (or higher) under the null hypothesis $\mathcal{H}_0 : \Lambda = 0$. If the $p$-value is below the significance level, then we deduce that the measured \ac{LR} $\hat{\Lambda}$ is significant. A false positive occurs when the $p$-value is below the significance level $\alpha$ but the null hypothesis $\mathcal{H}_0$ is true, i.e., there is no actual dependence between the variables $\Lambda = 0$, yet $\hat{\Lambda}$ is considered significant. Ideally, we expect the proportion of false positives (the \ac{FPR}) to match the significance level $\alpha$, i.e., one would expect an \ac{FPR} of 0.05 for a 5\% significance level $\alpha$. This same procedure is used for all linear dependence measures in this paper, with differing statistics and null distributions.

When the FPR is measured over $R$ trials (usually $R=1\ 000$ in this paper), confidence intervals can be determined based on the binomial distribution of $R$ draws of a random variable each with $\alpha=0.05$ chance of success.

\section{Experimental setup}

\subsection{Numerical simulations}
\label{app:numerical-simulations}

For our experimental validation, we use an \ac{AR} model similar to the example proposed in~\cite{barnett2011behaviour}, with two processes $\mb{X}$ and $\mb{Y}$ that have no interdependence, digitally filtered to increase their autocorrelation.
We begin with the $m$-variate time-series data $\mb{z} = (\mb{z}(1),\ldots,\mb{z}(T))$, i.e., an $m\times T$ matrix, where each realisation is generated from a first-order vector \ac{AR} model:
\begin{equation} \label{eq:mv-ar}
\mb{z}(t) = \mb{\Phi}(1) \mb{z}(t-1) + \mb{a}(t),
\end{equation}
with
\begin{equation} \label{eq:mv-model}
\mb{z}(t) = \begin{bmatrix}
\mb{x}(t) \\ \mb{y}(t)
\end{bmatrix}, \hspace{10px}
\mb{\Phi}(1) =
\begin{bmatrix}
\mb{\Phi}_{\mb{X}}(1) & \mb{0} \\
\mb{0} & \mb{\Phi}_{\mb{Y}}(1) \\
\end{bmatrix},
\end{equation}
and \ac{AR} parameters $\mb{\Phi}_{\mb{X}}(1) = 0.3 \mb{I}_k$ and $\mb{\Phi}_{\mb{Y}}(1) = -0.8 \mb{I}_l$.
The innovations $\mb{a}(t) = ( a_{1}(t),\ldots,a_{m}(t) )'$ are uncorrelated with mean $\mb{0}$ and unit variance matrix $\V{\mb{a}(t)} = \mb{I}_m$.
The matrix $\mb{z}$ is then partitioned into $k \times T$ and $l \times T$ matrices denoted $\mb{x}(t)$ and $\mb{y}(t)$. For each measure, we are interested in either the mutual information between $\mb{x}$ and $\mb{y}$, or the Granger causality from $\mb{y}$ to $\mb{x}$. If a third (conditional) process $\mb{w}$ is required to contextualise these measures (in their conditional forms), we consider another first-order \ac{AR} process $\mb{w}$ again using Eq.~\eqref{eq:mv-ar}, with $\mb{w}(t) \in \mathbb{R}^c$, $\mb{\Phi}_{\mb{W}}(1) = 0.4 \mb{I}_{c}$ and unit variance innovation process.
Each process, $\mb{x}$, $\mb{y}$, and $\mb{w}$, are then independently filtered with either an \ac{FIR} or an \ac{IIR} filter.
Both filters were low-pass, with their cutoff set to a normalised frequency of $\pi/2$ radians and a variable filter order. 

In our study of the mutual information between bivariate processes, we inject a causal influence from the univariate process $Y$ to $X$ in order to test the \ac{TPR} (i.e., statistical power of the test).
In this scenario, we generate bivariate time-series data $\mb{z}$ from same state equations (Eq.~\eqref{eq:mv-ar}) but with a small causal influence from $Y$ to $X$ in the \ac{AR} parameters:
\begin{equation}
\mb{\Phi}(1) = \begin{bmatrix}
\Phi_{X}(1) & \Phi_{XY}(1) \\
0 & \Phi_{Y}(1)
\end{bmatrix},
\end{equation}
with $\Phi_{XY}(1) = 0.03$, whilst $\Phi_{X}(1) = 0.3$ and $\Phi_{Y}(1) = -0.8$, as per Eq.~\eqref{eq:mv-model}.
  
  \subsection{Human Connectome Project}
  \label{app:fmri}
  
  The HCP rsfMRI dataset~\cite{van2012human} comprises 500  subjects imaged at a 0.72 s sampling rate for 15 minutes in the (relatively quiescent) resting state. This results in 1\,200 observations of spatially dense time series data, which is then parcellated into 333 regions of interest in the brain. Thus, the dataset contains 500 subjects with 333 brain regions each, and each of these regions is associated with a stationary time series of 1200 observations. The raw (BOLD) data of each region was then preprocessed by removing the DC component, detrending, applying a 3rd order zero-phase (or forward and reverse) Butterworth bandpass filter (0.01--0.08 Hz). These are common techniques used to remove potential artefacts. We also removed 200 observations from the start and the end of the time series, in order to minimise filter initialisation effects. This leaves $T=800$ observations for the analysis. In order to build a scenario where the null hypothesis holds, we conduct experiments on 1\,000 time-series pairs, selecting different random regions of interest from different random subjects, making the corresponding time series completely independent of one another. The analysis was performed for mutual information (between both univariate and bivariate time series) and Granger causality in the same way as discussed for the simulated time-series experiments above. We use the same hypothesis testing procedure (discussed above) with a 5\% significance level for the $\chi^2$-test and our newly proposed modified $\Lambda$-test.

\section{Additional prewhitening tests}
\label{sec:apx-prewhitening}

\begin{figure}[t]
		\centering
		\subfloat[]{
			\includegraphics[width=\columnwidth]{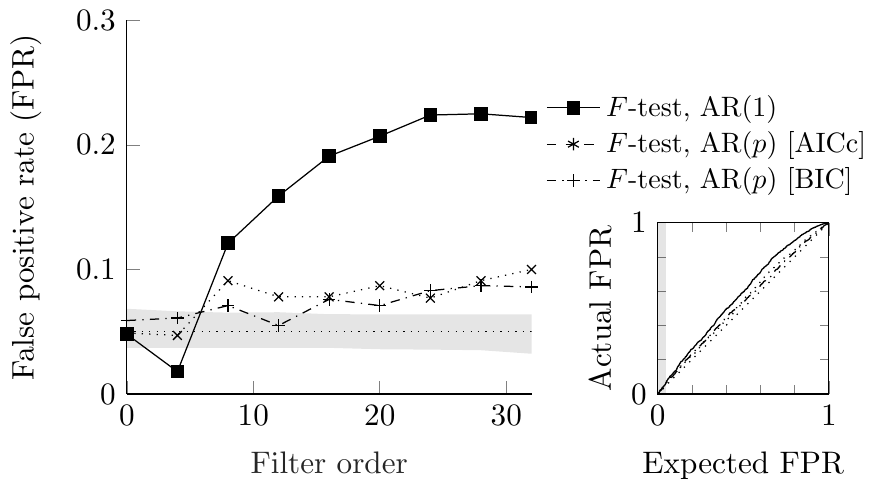}
			\label{subfig:mi-prewhitening-aicbic-fir}
		}
		\\
		\subfloat[]{
			\includegraphics[width=\columnwidth]{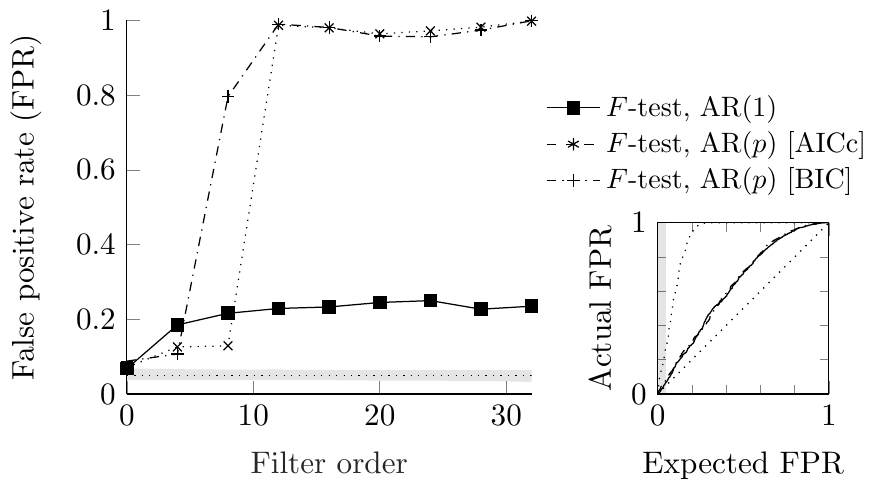}
			\label{subfig:mi-prewhitening-aicbic-iir}
		}
	\caption{
		Inferring the \ac{AR} models via AICc and BIC does not consistently improve prewhitening results for mutual information tests with FIR (\ref{subfig:mi-prewhitening-aicbic-fir}) or IIR (\ref{subfig:mi-prewhitening-aicbic-fir}) filtering.
		The experiments are as per \fig{mi-tests}, with two additional prewhitening approaches used: the $\text{AR}(p)$ model inferred from the AICc and BIC scores, respectively.}
	\label{fig:mi-prewhitening-aicbic}
\end{figure}

\begin{figure}[t]
		\centering
		\subfloat[]{
			\includegraphics[width=\columnwidth]{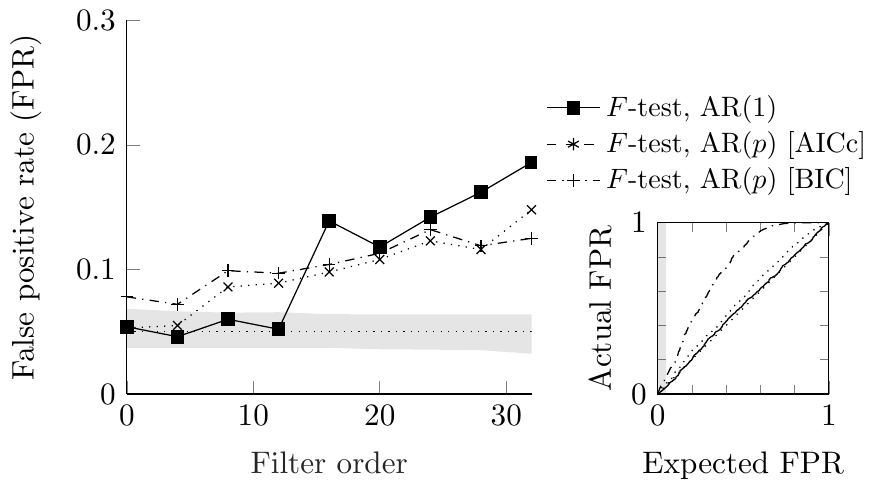}
			\label{subfig:gc-prewhitening-aicbic-fir}
		}
		\\
		\subfloat[]{
			\includegraphics[width=\columnwidth]{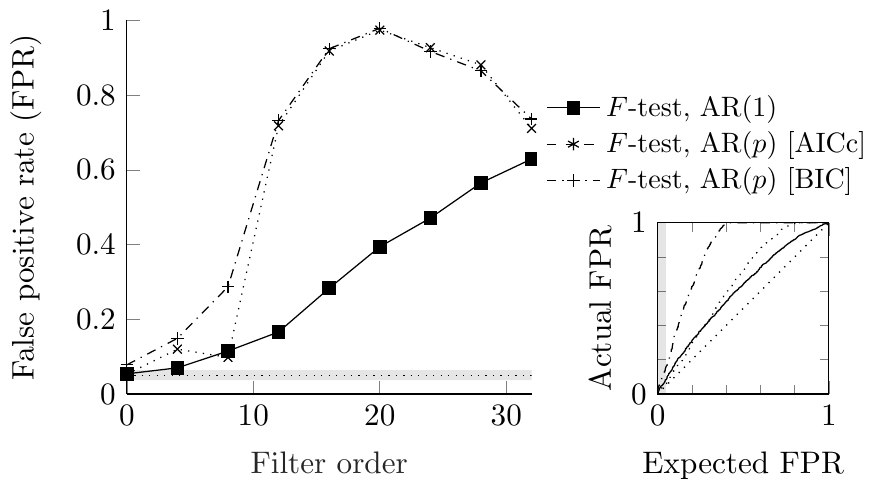}
			\label{subfig:gc-prewhitening-aicbic-iir}
		}
	\caption{
		Inferring the \ac{AR} models via AICc and BIC does not improve prewhitening results for Granger causality tests  with FIR (\ref{subfig:gc-prewhitening-aicbic-fir}) or IIR (\ref{subfig:gc-prewhitening-aicbic-fir}) filtering.
		The experiments are as per \fig{gc-tests}, with two additional prewhitening approaches used: the $\text{AR}(p)$ model inferred from the AICc and BIC scores, respectively.}
	\label{fig:gc-prewhitening-aicbic}
\end{figure}

This appendix provides further evidence that standard prewhitening techniques are insufficient for many covariance-stationary time series.
In the main text, we show that prewhitening time series is ineffective when the time-series models are either: AR($p$) models that are inferred from Burg's method; or ARMA($p$,$q$) models that are inferred from the BIC score (up to a maximum order of $p=q=5$).
Here, we extend these results to show that AR($p$) models inferred via the AICc (AIC with small-sample correction) and BIC scoring functions are also insufficient.

In \fig{mi-prewhitening-aicbic}, we show the extended prewhitening results for mutual information tests.
The algorithm iterates through all potential AR($p$) models and selects the one that minimises the AICc and BIC scores (independently).
For the FIR-filtered data, this approach shows a slightly increased \ac{FPR} above the nominal value of 5\%---this illustrates a marginal improvement over Burg's method from \fig{mi-prewhitening}.
However, for the IIR-filtered data, the \ac{FPR} approaches 100\% and is significantly worse than even methods with no or minimal prewhitening (c.f. the AR(1) models or the standard $F$-test in \fig{mi-tests}).
Similarly, prewhitening is shown to be insufficient for Granger causality tests in \fig{gc-prewhitening-aicbic}, where equivalent experiments were performed with no major qualitative differences (compared to \fig{gc-prewhitening}).

We do not show any further results for ARMA($p$,$q$) models, e.g., by increasing the maximum order or via a different criterion because the procedure to learn the parameters of higher-order ARMA models was too computationally expensive using the standard functions.
Given this constraint, all information criteria (AIC, AICc, or BIC) often chose the maximum order in practice, and so using alternative approaches was redundant.
We conjecture that prewhitening with ARMA($p$,$q$) models may perform favourably to AR($p$) models given the ability to infer arbitrarily complex models.
However, the constraints governed by the their inference procedures makes testing this currently intractable in practice.

\section{Considerations and extensions of Bartlett's formula} \label{sec:apx-mv-bartletts}

In our derivations we use a first-order approximation of Bartlett's formula (Eq.~\eqref{eq:bartletts-var}), that was originally described for spatially autocorrelated processes. However, since Bartlett's seminal work~\cite{bartlett1935some}, there have been a number of other extensions made to his formula as well as techniques intended to overcome the issues of its empirical computation.

One of the more general cases of Bartlett's formula is due to Roy~\cite{roy1989asymptotic}, who provided the large-sample distribution between pairs of sample cross-correlations at differing lags. Consider the four processes $Z_i, Z_j, Z_k, Z_l$.
Let
\begin{equation} \label{eq:delta}
\Delta_v(i,j,k,l) = \sum_{u=-\infty}^{\infty} \rho_{ij}(u) \rho_{kl}(u+v),
\end{equation}
where $\rho_{ij}(u)$ are the cross-correlation as per Eq.~\eqref{eq:corr}, and
\begin{equation}
	s_{ab}(v) = T^{\nicefrac{1}{2}} \, [r_{ab}(v) - \rho_{ab}(v)],
\end{equation}
as the standard error, with $r_{ab}(v)$ the sample cross-correlation in Eq.~\eqref{eq:inf-corr}.
In general, the asymptotic distribution of the standard error, $s_{ab}(v)$, is Gaussian with zero mean and covariance
\begin{align}\label{eq:bartletts-mv}
\lim_{T\to \infty} \Cov{s_{ab}(v)}{s_{de}(w)} \hspace{-90pt}& \nonumber \\
\approx& \, \Delta_{w-v}(a,d,b,e) + \Delta_{w+v}(b,d,a,e) \nonumber \\
&- \rho_{ab}(v) \big[\Delta_w(a,d,a,e) + \Delta_w(b,d,b,e)\big] \nonumber \\
&- \rho_{de}(w) \big[\Delta_v(b,d,a,d) + \Delta_v(b,e,a,e)\big] \nonumber \\
&+ \tfrac{1}{2} \rho_{ab}(w) \rho_{de}(w) \big[\Delta_0(a,d,a,d) + \Delta_0(a,e,a,e) \nonumber \\
&\hspace{78pt}+\Delta_0(b,d,b,d) + \Delta_0(b,e,b,e)\big]
\end{align}
This derivation can be further generalised to the non-Gaussian case, for instance by allowing for skewed distributions~\cite{su2012multivariate}. 
More recently, a first-order approximation of this formula was given by Afyouni et al.~\cite{afyouni2019effective}, which takes the same form as Eq.~\eqref{eq:bartletts-mv}, with Eq.~\eqref{eq:delta} slightly modified.

One consequence of knowing the full covariance structure (Eq.~\eqref{eq:bartletts-mv}) is that such a distribution could further help in situations when the partial correlation terms that Wilks' statistic decomposes into are themselves correlated.
That is, in the main text, we provided the variance for each $\Lambda^*$-distributed variable $L_i$ (as a beta distribution by assuming independence).
By assuming independence, we were able to obtain the sampling distribution of Wilks' criterion as a product of these $\Lambda^*$-distributions.
However, if the random variables were correlated, then we must use the multivariate form of Bartlett's formula~\eqref{eq:bartletts-mv}, which provides the covariance of each variable $L_i$ term with all other variables $L_j$ (i.e., we would have $n_{ij}$ for each $i$ and $j$, rather than just $n_i$).
Accounting for this covariance would require knowing the distribution of the general product of correlated beta-distributions that, to the best of our knowledge, is not an established result.

Another use-case of Eq.~\eqref{eq:bartletts-mv} is testing against an alternative hypothesis ($\mathcal{H}_1: \rho_{ab}(0) \neq 0$), which is discussed at length by Afyouni et al.~\cite{afyouni2019effective} for correlation coefficients.
One could follow the same logic from this paper to provide the alternative hypothesis test for linear dependence measures based on Wilks' statistic.

There are a number of special cases of Roy's formula that are worth noting. In the event that we are interested in the covariance $\Cov{s_{ab}(v)}{s_{ab}(w)}$ between cross-correlation estimates of two univariate processes $Z_a$ and $Z_b$ at arbitrary lags $v$ and $w$, this is obtained from Eq.~\eqref{eq:bartletts-mv} by setting $d = a$ and $e = b$, reducing to the results reported in~\cite{bartlett955introduction} and~\cite{brockwell1991time} (Theorem 11.2.3). Using this special case, the null distribution of Pearson (zero-lag) correlation between two univariate processes $\V{s_{ab}(0)}$ can be obtained by setting $v = w = 0$. Finally, under the assumption that $\rho_{ab}(0) = 0$, most of these terms disappear and we are left with Bartlett's original formula~\cite{bartlett1935some}:
\begin{align}
	\lim_{T\to \infty} \V{s_{ab}(0)} &\approx \lim_{T\to \infty} \V{T^{\nicefrac{1}{2}} \, [r_{ab}(0)]}, \nonumber \\
	&= \sum_{u=-\infty}^{\infty} \rho_{ii}(u) \rho_{jj}(u)
\end{align}
with a similar form given by a first-order approximation~\cite{bayley1946effective}:
\begin{equation} \label{eq:final}
	\V{s_{ab}(0)} \approx T^{-1} \sum_{u=-\infty}^{\infty} (T-|u|) \rho_{aa}(u) \rho_{bb}(u).
\end{equation}
Due to symmetry of the autocorrelation function about lag-zero for stationary processes, we can simply sum over the positive lags in Eq.~\eqref{eq:final}, $u>0$, which was the form used throughout this paper (see Eq.~\eqref{eq:bartletts-var}).
For the exact relationship between the large-sample approximations and the first-order approximations, we refer the reader to the discussions in~\cite{bartlett1946theoretical,quenouille1947notes,bayley1946effective}.
Bartlett did indeed present a formula irrespective of sample size~\cite{bartlett1946theoretical}, which may yield an improvement for small sample distributions and give minor practical advantages, however, we did not find this necessary for any experiments and instead follow the approximations in Eq.~\eqref{eq:final}.


Another potential source of error in the sampling distributions come from incorrectly estimating the autocorrelation function $r_{aa}(u)$. Tapering (also known as data windowing) is commonly used in practice to regularise the autocorrelation samples to better estimate their true value~\cite{chatfield2003analysis,afyouni2019effective}. These approaches involve scaling the autocorrelation samples by some factor, with the maximum lag truncated below the dataset length. Using this method, we can appropriate Bartlett's formula to
\begin{equation}
	\V{s_{ab}(0)} \approx 1 + 2\,  \sum_{u=1}^U \frac{T-u}{T} \lambda(u) r_{aa}(u) r_{bb}(u),
\end{equation}
where $\lambda(u)$ are a set of weights called the lag window and $U<T$ is the truncation point. The lag window comprises $\lambda(u)$ values that decrease with increasing $u$; two common approaches are the Parzen and the Tukey windows (see~\cite{chatfield2003analysis} for details). Numerous truncation points $U$ have also been proposed, e.g., $T/4$, $T/5$, $\sqrt{T}$, and $2\sqrt{T}$~\cite{afyouni2019effective}.


In the above few sections we outlined a number of potential factors that could introduce small size or power distortions in our hypothesis tests. To compare our approach with these more complex extensions, we ran experiments with the effective sample size computed from the full covariance matrix~\eqref{eq:bartletts-mv}, both with and without tapering.
These were computed for the experiments from Fig.~\ref{fig:mi-tests} in the paper, however, as mentioned above, the product of correlated beta- or $F$-distributed variates is unknown and so our modifed $\Lambda$-test could not be performed.
Instead, we used the sums of $z$-transformed partial correlations (each of which make up the conditional mutual information term), rather than the products of squared partial correlations.
That is, by transforming each partial correlation, we expect the sum of these correlation to be approximately Gaussian.
In performing these tests, we found no notable difference in the \acp{FPR} for any of the validation experiments, suggesting that these additions made no significant difference towards reducing size or power distortions.

\section{Partial autocorrelation and active information storage} \label{sec:apx-ais}

The partial autocorrelation function conveys important information regarding the dependence structure of an \ac{AR} process~\cite{brockwell1991time}.
For a univariate stationary time series $Z$, the partial autocorrelation $\alpha_Z(u)$ at lag $u$ is the correlation between $Z(t)$ and $Z(t-u)$, adjusted for the intervening observations $\mb{Z}^{(u-1)}(t) = [Z(t-1) \ssep \ldots \ssep Z(t-u+1)]$.
Denote $Z^u$ as the process of $Z$ lagged by $u$ time steps and $Z^{(u)}$ as the history up until that lag (inclusive) $\mb{Z}^{(u)} = [Z^1 \ssep \ldots \ssep Z^u]$.
Then, for a stationary time series, the partial autocorrelation function is defined by~\cite{brockwell1991time}
\begin{equation}
\alpha_Z(1) = \rho_{ZZ^1},
\end{equation}
and
\begin{equation}
\alpha_Z(u) = \rho_{ZZ^{u} \cdot \mb{Z}^{(u-1)} }, \hspace{10px} u > 1.
\end{equation}
Although we use Burg's method to identify the relevant history length $p$ for an \ac{AR} model of $Z$ length for AR models in this paper, it is common practice to use the partial autocorrelation function instead, since $\alpha_Z(u) = 0$ for $u > p$~\cite{brockwell1991time,reinsel2003elements,box2015time}. Again, this is a statistical estimate and thus the order $p$ is inferred by testing each sample partial autocorrelation $\hat{\alpha}_Z(u)$ for significance against the null distribution.

Intriguingly, our work reveals a relationship between the partial autocorrelation function and active information storage~\cite{lizier2012local}---a recently developed model-free measure for quantifying memory in a process---under the linear Gaussian assumption.
The average active information storage $\mathcal{A}_{X}$ quantifies the information storage in a process.
For a $p$-order Markov process $X$, this is quantified by the mutual information between the relevant history $\mb{X}^{(p)}(t)$ and variable $X(t)$, i.e.,
\begin{equation} \label{eq:ais}
\mathcal{A}_{X}(p)  = \mathcal{I}_{X ; \mb{X}^{(p)}}.
\end{equation}
Since the average active information storage is a specific type of mutual information, we can use the chain rule to decompose it into a sum of squared partial autocorrelations:
\begin{align} \label{eq:ais-as-rho}
\mathcal{I}_{X ; \mb{X}^{(p)}} &= - 1/2 \sum_{u=1}^{p} \log \left( 1 - \rho_{X X^u \cdot \mb{X}^{(u-1)} }^2 \right), \nonumber \\
&= - 1/2 \sum_{u=1}^{p} \log \left( 1 - [\alpha_X(u)]^2 \right).
\end{align}
This same logic can be straightforwardly applied to other measures such as excess entropy~\cite{bialek2001complexity} and predictive information~\cite{crutchfield2003regularities}.

In additional to quantifying the memory within a process, active information storage is often used for inferring the optimal history length for both the Gaussian and non-Gaussian cases~\cite{lizier2014jidt}.
This is typically achieved by using the $\chi^2$-test to infer the significance of increasing the embedding lengths $p$.
For AR processes with Gaussian innovations, we infer the embedding length $p$ for $X$ by first taking the difference $\delta_x(u) = \mathcal{A}_x(u+1) - \mathcal{A}_X(u)$ and then generating a $p$-value by testing $2\, \delta_X(u)$ against a $\chi^2$(1) distribution, which represents the null hypothesis of no increase in information storage.
If the $p$-value is below a threshold (say 5\%), then the test is rejected and the lag is increased $u = u + 1$.
This process is iterated until the null hypothesis is accepted, at which point we surmise that the optimal lag $p$ is the one at which $\delta_X(p+1)$ is considered insignificant. This approach is similar to using the partial autocorrelation, as the difference $\delta_X(u)$ is equivalent to squared partial autocorrelation up to a factor of two. This can be seen from Eq.~\eqref{eq:ais-as-rho}:
\begin{align}
\delta_X(u) &= \mathcal{A}_X(u+1) - \mathcal{A}_X(u) \nonumber \\
&= -\frac{1}{2}\log \left( 1 - [\alpha_X(u+1)]^2 \right)
\end{align}
In contrast to measures of dependence between multiple processes, the $\chi^2$-test appears suitable here (without adjusting for an effective sample size) for testing $\delta_X(u)$ for $u>p$.
This is because, after the full set of past variables is included in the regression, any higher order residuals $x^{u} - \hat{x}^{u}(\hat{x}^{(p)})$ with $u > p$ have statistically zero autocorrelation for every lag.



\begin{acknowledgments}
	
	OMC and BDF were supported by NHMRC Ideas Grant 1183280.
	JTL was supported through the Australian Research Council DECRA grant DE160100630.
	JMS was supported through a University of Sydney Robinson Fellowship and NHMRC Project Grant 1156536.
	JMS and JTL were supported through The University of Sydney Research Accelerator (SOAR) Fellowship program.
	OMC, JTL and JMS were supported through The Centre for Translational Data Science at The University of Sydney's Research Incubator Funding Scheme.
	OMC, JTL, JMS and BDF were supported through the Centre for Complex Systems at The University of Sydney's Emerging Aspirations scheme.
	High performance computing facilities provided by The University of Sydney (Artemis) have contributed to the research results reported within this paper.
\end{acknowledgments}

\end{document}